\documentclass[useAMS,usenatbib]{mn2e}
\usepackage{graphics,epsfig,psfig}
\usepackage[normalem]{ulem}
\usepackage{xcolor,verbatim}
\usepackage[]{inputenc,amssymb, amsmath}

\def \be{\begin{equation}}
\def \ee{\end{equation}}
\newcommand       \ba           {\begin{eqnarray}}
\newcommand       \ea           {\end{eqnarray}}
\def \bea{\begin{eqnarray}}
\def \eea{\end{eqnarray}}

\newcommand{\comments}[1]{}

\newcommand       \apjl         {ApJL}
\newcommand       \apjs         {ApJS}
\newcommand       \aap          {A\&A}

\definecolor{webgreen}{rgb}{0,.5,0}
\definecolor{webbrown}{rgb}{.6,0,0}
\usepackage[pdfpagelabels]{hyperref}
\hypersetup{%
   colorlinks=true,%
   breaklinks=true,%
   plainpages=false, bookmarksnumbered, bookmarksopen=true,
   bookmarksopenlevel=1,%
   urlcolor=webbrown, linkcolor=blue, citecolor=webgreen,
   }

\title[A simple model for gas in CGM and ICM]{A 1-dimensional hydrodynamic model for accretion, cooling and heating of gas in dark matter halos from $z=6$ to $z=0$}  

\author[P. P. Choudhury, G. Kauffmann, P. Sharma]
{Prakriti Pal Choudhury$^\dag$$^\ddag$, Guinevere Kauffmann$^\ddag$, Prateek Sharma$^\S$ \\
$^\dag$ Department of Physics, Indian Institute of Science, Bangalore 560012 , India (prakritic@iisc.ac.in)\\
$^\ddag$Max Planck Institute for Astrophysics, Garching 85748, Germany (gamk@mpa-garching.mpg.de) \\
$^\S$Department of Physics and Joint Astronomy Program, Indian Institute of Science, Bangalore, India 560012 (prateek@iisc.ac.in)}

\begin{document}
\maketitle

\label{firstpage}
\begin{abstract}
We study an idealized 1D model for the evolution of hot gas in dark matter halos for redshifts $z=[0,6]$. We introduce a numerical setup incorporating cosmological accretion of  gas, along with the growth of the halo,
based  on the Van den Bosch model for the average growth of halos as a function of cosmic time. We evolve one-dimensional Lagrangian shells 
with radiative cooling of the gas and heating due to feedback from the gas  
cooling and moving in toward the center. 
A simple Bondi accretion model on to a central black hole is used to include feedback heating. 
The setup captures some of the key characteristics of spherically symmetric accretion onto the halos: 
formation of virial shocks slightly outside $r_{200}$ and long-term thermal balance in the form of cooling and heating cycles. 
The gas density outside our initial halos at $z=6$ is constrained by requiring that the baryon fraction within the virial radius  for non-radiative evolution be equal to the universal value at almost all times. The 
total mass in the cold phase (taken to be $\sim 10^4$ K) within $40$ kpc is tracked as a function of the halo 
mass and redshift. We compare the evolution of the cold gas mass to the 
observed stellar-mass 
versus halo mass relations, following which, we can constrain the feedback energy required
for different halo masses and redshifts. 
We also compare and match the hot gas density and temperature profiles for our most massive halo to those of clusters observed upto redshift $2$. 
Our model is thus an improvement over the semi-analytic models in which isothermal condition and $\rho \propto r^{-2}$ are assumed. 
\end{abstract}
\begin{keywords}
galaxies: halos -- galaxies: cooling flows -- methods: numerical. 
\end{keywords}

\section{Introduction}
Galaxy formation is known to be an outcome of gravitational processes (clustering) and gas-dynamical effects acting together. The success of N-body simulations  
lies in their ability to predict the detailed clustering properties of galaxies from initial conditions set by the $\Lambda$CDM cosmological model (\citealt{1985Nature_frenk}, \citealt{1985Apj_davis}, \citealt{2000MNRAS_colberg}). However, radiative cooling and 
dissipation become important when attempting to predict the detailed physical properties of galaxies, such as their stellar masses and star formation rates. Semi-analytic models with radiative cooling, star formation and 
quenching by AGN and supernova feedback, provide a simple description of the relevant gas physics, but are unable to track the spatial distribution of the gas in any detail (\citealt{2005Nature_sw}, \citealt{2006MNRAS_croton}).

The advent of cosmological hydrodynamic simulations (\citealt{1989ApJ_hernquist}, \citealt{2000gadget2}, \citealt{2001gadget}, \citealt{2002ApJ_murali}, \citealt{2014MNRAS_illustris}) provided  better means to treat gas dynamics more self-consistently. However, current simulations are very expensive or do not have the sufficient mass resolution to study the small scale physics in detail in very large cosmological volumes (\citealt{2018MNRAS_pillepich}, \citealt{2017MNRAS_tremmel}). In addition, ``sub-grid'' models are used for studying small-scale, unresolved physical processes like star formation and feedback (\citealt{2002MNRAS_Kay}, \citealt{2005Nature_matteo}, \citealt{2008ApJ_puchwein}, \citealt{2015MNRAS_crain}). Three dimensional hydrodynamic simulations have been able to track the spatial distribution of gas with reasonable accuracy, but it is challenging to study the role of different subgrid recipes in a controlled way. Compared to these, one-dimensional hydrodynamic models are computationally less expensive and often provide valuable insight into the physics of gas and its radial distribution, while retaining simplicity.  It is easy to generalize these to more realistic 2D/3D  simulations and perform controlled parameter survey to elucidate the physical inputs that are necessary to match different aspects of the observational data.

In the standard picture of spherical infall of gas (\citealt{1972ApJ_gunngott}, \citealt{1985ApJS_bertschinger}) into the dark matter halos, the accreting IGM is heated to the 
halo virial temperature behind an expanding virial shock. The gas is first supported by pressure in quasi-hydrostatic equilibrium. It cools radiatively, gradually contracting and forming a cold disk in which star formation happens (\citealt{1980MNRAS_fall}, \citealt{1998MNRAS_momaowhite}). One-dimensional Lagrangian models of gas in dark matter halos, incorporating this standard picture,  have been around for a long time.  
One of the earliest papers (\citealt{1988MNRAS_thomas}) studies multiphase gas in the intracluster medium with balance of cooling and inflow. However, only gas is evolved in this model and not dark matter.  

\cite{thoul_weinberg1_94} introduced a spherical collapse of two-fluid system including gas (radiatively cooling) and dark matter. 
In our model, we use a numerical scheme similar to \cite{thoul_weinberg1_94} but with different treatment of gravity due to dark matter, radiative cooling and heating that is expected from supernovae/AGNs. This combined model of cosmological growth of halos and realistic energetics of gas has not been carried out before in a simple, Lagrangian set-up. The first treatment of gravity as a gradually deepening potential well, that we use in our model, was presented in \cite{1978ApJ_parrenod}. \cite{1997MNRAS_knight} later added radiative cooling and concluded that in low-mass halos the baryon content is overpredicted in such a model. Thus the idea of a central source of heating became indispensable. 

Through a simple spherical 1D model which incorporates dark matter and gas, \citealt{2003MNRAS_birnboimdekel} (see also \citealt{1991ApJ...379...52W}) showed that the virial shock in smaller halos ($\lesssim 10^{12}~{\rm M_{\odot}}$), is not stable and in such cases gas is not heated to the virial temperature due to efficient radiative cooling and directly falls to the halo center (this is known as cold-mode accretion).\footnote{Even low mass halos may have hot gas because of feedback and mergers (e.g., see \citealt{sokolowska_2017}). Earlier works on cold mode only considered radiative cooling and ignored heating due to mechanical feedback.} In 3-D cosmological simulations the cold mode appears in the form of a number of cold streams feeding the central galaxy (\citealt{2005MNRAS_keres}, \citealt{2009Nature_dekel}, \citealt{2009MNRAS_Keres}; but also see \citealt{2013MNRAS.429.3353N}); however, cold streams cannot be captured by a 1D model such as ours.  The virial shock is very stable at cluster scales and the hot gas is close to hydrostatic equilibrium (\citealt{2018arXiv_oppen}), except in the central parts where the cooling time is usually shorter. The gas that cools out within the central region of the massive halos, is considered to be the only source of fuel for star formation.


In order to quantify stellar content in halos, the relation between the stellar mass of the central galaxy and the host halo mass has been parameterized extensively for a wide range of halos. The mass function of dark matter halos, calibrated from large-scale simulations, is combined with the observed number density of galaxies as a function of their stellar mass using a technique called statistical abundance matching (\citealt{2010ApJ_moster}, \citealt{2013ApJ_moster}). We adjust our very simple feedback model parameters to match the average stellar mass and halo mass relation. While bigger halos have higher stellar mass, the ones hosting more massive black holes are observed to have smaller star formation rates on average (\citealt{2018Nature_navarro}). But this only indirectly suggests that energy feedback from black hole may regulate cooling in large gaseous halos. The clearest evidence of black hole feedback comes from observations of synchrotron-emitting radio lobes that are co-spatial with the X-ray cavities found in galaxy clusters. The work done to inflate such huge bubbles, comparable to the energy required to prevent cooling flows (\citealt{fabian94}, \citealt{2011ApJ_cavagnolo}) in clusters, can only come from accretion onto supermassive black holes. 

Earlier detailed observations of gaseous halos study clusters and groups in X-rays at low redshifts. The radial profiles of density, temperature and entropy have been  extensively studied for low-redshift ($z=0.05-0.2$) clusters (\citealt{2005ApJ_vikhlinin}, \citealt{2005A&A_pointecouteau},\citealt{2006ApJ_vikhlinin}, \citealt{2006ApJ_kotov}, \citealt{accept09}). Quenching of the cooling of hot gas, has been seen in the observations of cavities that put important constraints on AGN outburst energy and mean jet power (\citealt{2004ApJ_birzan}, \citealt{2011ApJ_cavagnolo}, \citealt{1993MNRAS_boehringer}, \citealt{2003MNRAS_fabian}, \citealt{2000ApJ_mcnamara}). 
Recently the redshift evolution of galaxy clusters is traced for SZ selected clusters using the X-ray data (\citealt{2011ApJ_marriage}, \citealt{2014ApJ_mcdonald}, \citealt{2013ApJ_mcdonald}, \citealt{2017ApJ_nurgaliev}, \citealt{2017ApJ_mcdonald}). It hints at the invariant cooling properties and X-ray morphology of clusters till $z \approx 2$. We present comparisons of the radial gas density profiles of such SZ selected halos with that of the cluster in our model. Current observations also study the radial profiles of baryon fraction in clusters and groups by tracking the free electrons using kinetic SZ effect (e.g \citealt{2016PhRvD_schaan}) and may provide constraints on the simulations.  

In this paper, we propose a very simple yet useful 1-D model for the evolution of the gas in smoothly growing, isolated dark matter halos with only a few adjustable parameters. The model is tuned to match the observed relation between stellar mass and halo mass. We disregard direct effects of mergers as well as processes affecting satellite galaxies, such as ram pressure stripping and strangulation. While this is a limitation, this allows us to focus on the interplay of smooth build-up of the halo and feedback heating and cooling. 
From our models, we get the following important perspectives on the state of baryons within the halos: (1) the radial profiles of gas density and temperature in massive clusters, 
(2) total feedback energy required to regulate cooling and heating in halos of different masses (3) how the baryon fraction evolves when we constrain the global energetics using the relation of stellar mass and halo mass. This model can be useful to modify the semi-analytic models of galaxy formation that do not follow gas-dynamical processes in detail. This model may also provide an useful middle-ground between the idealized simulations of isolated clusters, groups and smaller halos (\citealt{2017MNRAS_fielding}) that do not evolve the halo with redshift and the cosmological hydrodynamic simulations. The latter, besides being expensive, often lack sufficient resolution to study the diffuse gas (ICM/CGM) away from the center in an individual halo. Hence future extension of our model to 2D/3D will be relevant for quantifying the content of multiphase gas in the outskirts of the halos, which will be particularly interesting with respect to the COS-HST survey (\citealt{tumlinson17}).

The paper is organized as follows. In section \ref{sec:physset} we describe the set-up for our numerical experiments including the initial conditions, in section \ref{sec:res} we present all our results for four different halos with a wide range of masses excluding and including radiative cooling and feedback and finally in section \ref{sec:disc} we discuss the implications of our results and caveats in detail. 

\section{Physical set-up}
\label{sec:physset}
In this paper we evolve four halos namely $M_{200,z=0} = 5 \times 10^{14}M_{\odot} (M_{14}), 5 \times 10^{13}M_{\odot} (M_{13}), 5 \times 10^{12}M_{\odot} (M_{12}), 5 \times 10^{11}M_{\odot} (M_{11}) $ from $z=6$ to $z=0$. The dark matter halo does not evolve dynamically. We use a parametric form for dark matter density profile and vary the parameters as functions of time/redshift. The gas, however, is evolved hydrodynamically. At all times the halos have an inner radius at $r_{\rm min} = 1~{\rm kpc}$. The inner boundary of the innermost shell is always pined to $r_{\rm min}$ and the velocity is set to zero there. For runs with radiative cooling, we include a careful treatment of the gas shells that move too close to this innermost trajectory (described in section \ref{sec:freez}).

\subsection{Evolution of dark matter halos}
We do not evolve the equation of motion for the dark matter shells and simply approximate the increase in the gravitational potential due to the growth of the halo using the fitted average mass accretion histories proposed by \cite{VDB2002}. These authors use extended Press-Schechter formalism (\citealt{1974ApJ_pressschechter}) to show that the average mass accretion history follows a simple universal form for a wide range of halos and cosmologies. They then follow the most massive progenitor of the halos at the current redshift,  back in time, in the Millennium simulation (\citealt{2005Nature_sw}) and compare their analytical formulae with the mass accretion histories derived for halos in the Millennium simulation. \footnote{They find reasonable agreement and only mild inconsistencies which could be easily interpreted as a consequence of ellipsoidal collapse in the simulations as opposed to the assumption of spherical collapse in the formalism.} In this paper we adopt the recipe that calculates the analytical formulae.The mass accretion history or $M_{200}(t)$ is evaluated using Eq. (A1) in the APPENDIX A of \cite{VDB2002} in which we use the parameters ($z_f$ and $\nu$) as described in the same Appendix for $\Lambda$CDM cosmology. 

\begin{figure}
\centering
 \includegraphics[width=0.5\textwidth]{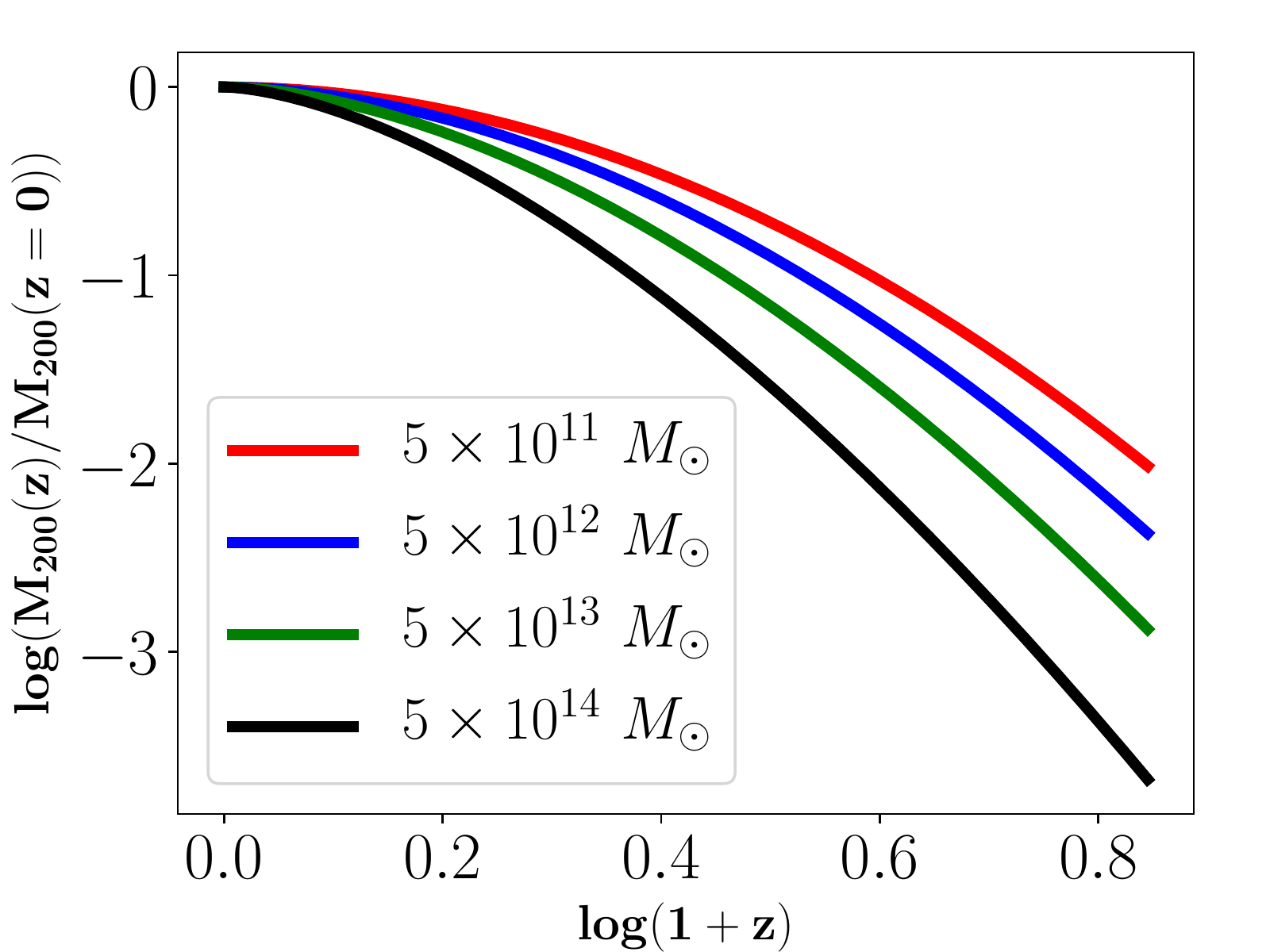}
 \caption{The evolution of halo mass for our four different halos, corresponding to the average mass accretion histories considered with $\Omega_{\Lambda} = 0.75$, $\Omega_{\rm m} = 0.25$. These profiles are very similar to the profiles in Figure 5 of \citealt{VDB2002}, which correspond to different values of the cosmological parameters $\Omega_{\Lambda}$ and $\Omega_{\rm m}$. Note that the massive halos accrete more mass (even when normalized to the halo mass) compared to the lower mass halos.}\label{fig:fig01} 
 \end{figure}

Figure \ref{fig:fig01} shows the halo mass as function of redshift in our halos. In order to evolve the halo masses with time, we use the following cosmological parameters, $\Omega_{\Lambda} = 0.75$, $\Omega_{\rm m} = 0.25$ and $\sigma_8 = 0.9$. The concentration parameter ($c$) is modelled using \cite{Zhao2009}, which takes into account that the halo concentration is tightly correlated with the time at which the main progenitor of a halo gains $4$\% of its final mass. It can be written in the following form (Eq. 13 in \citealt{Zhao2009}):
\ba
c = 4\Big(1 + \Big(\frac{t}{3.75~t_{0.04}}\Big)^{8.4}\Big)^{1/8},
\ea
where $t_{0.04}$ is the time when the main progenitor gained $4\%$ of its final mass.
We use a gravitational acceleration due to $M_{200}(t)$ of the following form (\citealt{nfw96}),
\ba
\label{eq:eq1}
g(r,t) &=& \frac{d\phi_{\rm NFW}}{dr} \\
\nonumber 
=\frac{GM_{200}(t)}{F(c(t,M_{200}))} \Big(\frac{\ln(1 + r/r_s(t))}{r^2} &-& \frac{1}{r(r_s(t) + r)}\Big),\\
\nonumber
\ea
where
\ba
\nonumber
F(c) = \ln(1+c) &-& \frac{c}{1+c}
\ea
and
\ba
r_s(t) = r_{200}(t)/c(t)
\ea
is the scale radius. 

Thus the entire information of the dark matter is encapsulated in the expression of the gravitational acceleration that becomes a function of time (or equivalently, redshift). The inherent assumption is that within each hydrodynamic step for evolving the gas, the dark matter comes to equilibrium instantaneously. We use the same gravitational acceleration beyond $r_{200}$. However, \citealt{diemer_kravt2014} suggest that dark matter density profile does not follow NFW very far away from the halo.
 \subsubsection{Effects of discrete mergers}
 The modelling of violent, discrete merger events is beyond the scope of this work. We have an average, smooth growth due to cosmological infall as well as mergers. Figure 3 of \cite{VDB2002} shows how the average behavior of the halo is extracted from the growth history of individual halos undergoing a number of mergers and this mean evolution is what we use.  Mergers have two important influences on the halo: the dark matter mass increases rapidly; the gas and the stars are redistributed. We do not include these effects but these can be incorporated in the 2D/3D generalisations of our model.

\subsection{Initial conditions for gas}
\label{sec:ini}
We start our runs at $z=6$ where we initialize the gas within the approximate virial radius, $r_{200}$ (the mean density within which is 200 times the critical density of the universe), to be in hydrostatic equilibrium confined by the gravity due to the dark matter halo. Outside $r_{200}$ we use a parametric density profile. This is the density profile that we set in the outskirts of the halo so that the mass accreted within the halo at each time can preserve the universal baryon fraction for non-radiative evolution. This parametric profile may not describe gas density profile outside the halo accurately but serves as a reservoir of gas that falls onto the halo over the entire range of redshifts. We will describe these two profiles (within and outside the halo) in the next two subsections (section \ref{sec:hse} and \ref{sec:o1}). The general density profile can be described as follows:
\ba
\rho(r) &=& \rho_{HSE}(r) ~ r \leq r_{200} \\
           &=& \rho_{O}(r) ~ r > r_{200},
\ea
where $\rho_{HSE}(r)$ is what we initialize inside $r_{200}$ and $\rho_{O}(r)$ is the density profile that we set outside $r_{200}$. The matching condition for the density profile inside and outside initially at $z=6$, is 
\ba
\label{eq:match}
\rho_{HSE}(r_{200}) = \rho_{O}(r_{200}).
\ea

We assume that initially the gas is isentropic throughout. This assumption breaks down inside the halo after an accretion shock is formed. With this initial density profile for all the gas shells we calculate pressure ($p$), temperature ($T$) and energy density per unit mass ($u$) in the following way:

\ba
p(r) &=& K_0 \rho(r)^{\gamma}, \\
T(r) &=& p(r)/n(r) k_B ,\\
u(r) &=& p(r)/(\gamma-1) \rho(r),
\ea

where $\gamma=5/3$, $k_B$ is the Boltzmann constant, $n(r) = \rho(r)/\mu m_p$ is the number density, $K_0 = K f_T k_B/(\mu m_p)(\mu_e m_p)^{\gamma -1}$ is an entropy index in which $f_T = 1.16 \times 10^7$ is the conversion factor of temperature from keV to kelvin(K). We have used the mean molecular weights, $\mu=0.62$ and $\mu_e=1.17$. We use the self-similar scaling of entropy $K = C\Big(\frac{M_{200}(z=6)}{M_0}\Big)^{\frac{2}{3}}~{\rm keV}~{\rm cm}^2$, where $C=2$ is a normalization constant that we fix using the halo $M_{13}$ that has mass $M_0$ at $z=6$. This scaling is motivated by the X-ray entropy $K=T_{\rm keV}/n_e^{2/3}$, popularly used in the galaxy cluster literature. The temperature of the IGM gas outside $r_{200}$ is much lower than the virial temperature because the density set by $\rho_{O}$ profile (see Figure \ref{fig:fig03}) is also low. Therefore, the gas falling onto the halo is supersonic and forms a virial shock close to $r_{200}$. 

We have tested that the evolution of the gas is independent of the exact temperature profile outside $r_{200}$ (as long as it is much smaller than the virial temperature).  All the shells (both inside and outside) have an initial Hubble-flow velocity $v_i = H(z=6)~r$ as given in section $3.2$ of \cite{thoul_weinberg1_94}  for the case of self-similar collapse of a collisional, non-radiative gas.

\subsubsection{Density inside $r_{200}$: $\rho_{HSE}$}
\label{sec:hse}
We calculate $\rho_{HSE}$ assuming hydrostatic equilibrium inside $r_{200}$. Therefore, we solve the following equations,
\ba
\label{eq:eq5}
\frac{dp}{dr} &=& -\rho g ,\\
\label{eq:eq6}
p &=& n k_B T , \\
\label{eq:eq7}
K &=& \frac{T_{keV}}{{n_e}^{\gamma -1}},
\ea
where $T_{keV} = T/f_T$, $g$ is calculated from Eq. \ref{eq:eq1} at the time $t(z=6)$ and all the other symbols are as described above. In order to solve these equations we need to assume a boundary condition for gas density at $r_{200}$. Since $r_{200}\propto {M_{200}}^{1/3}$, the outer density for all the halos can be assumed to be identical. We make sure here that the parameters $K$ (as discussed in the previous section) and density at $r_{200}$, are chosen such that the total amount of gas inside the halo at $z=6$ is close to the universal baryon fraction (within around $10-15\%$; Figure \ref{fig:fig04}). It is worth noting that this is only the initial boundary condition and the density will vary according to the hydrodynamic evolution for times $t > t_{z=6}$. The red curves in Figure \ref{fig:fig03} shows the initial number density and temperature for halos $M_{14}$ and $M_{11}$.

\subsubsection{Density outside $r_{200}$ : $\rho_{O}$}
\label{sec:o1}
For the initial outer ($>r_{200}$) gas density profile, we use a broken power-law. If density at $r_{200}$ is denoted as $\rho_{r_{200}}$ and the density at $2r_{200}$ as $\rho_{2r_{200}}$, the general form of the density power-law is chosen to be
\ba
\nonumber
\rho_{O}(r) &=& \rho_{r_{200}}{\Big(\frac{r}{r_{200}}\Big)}^{-4}~~r_{200} < r \le 2r_{200},\\
\label{eq:eq8}
\rho_{O}(r) &=& \rho_{2r_{200}}{\Big(\frac{r}{2r_{200}}\Big)}^{-\alpha}~~ r \ge 2r_{200},
\ea
where a different $\alpha$ is obtained for different halos (5th column in Table \ref{table:haloruns}). 

\begin{figure}
\centering
 \includegraphics[width=0.5\textwidth]{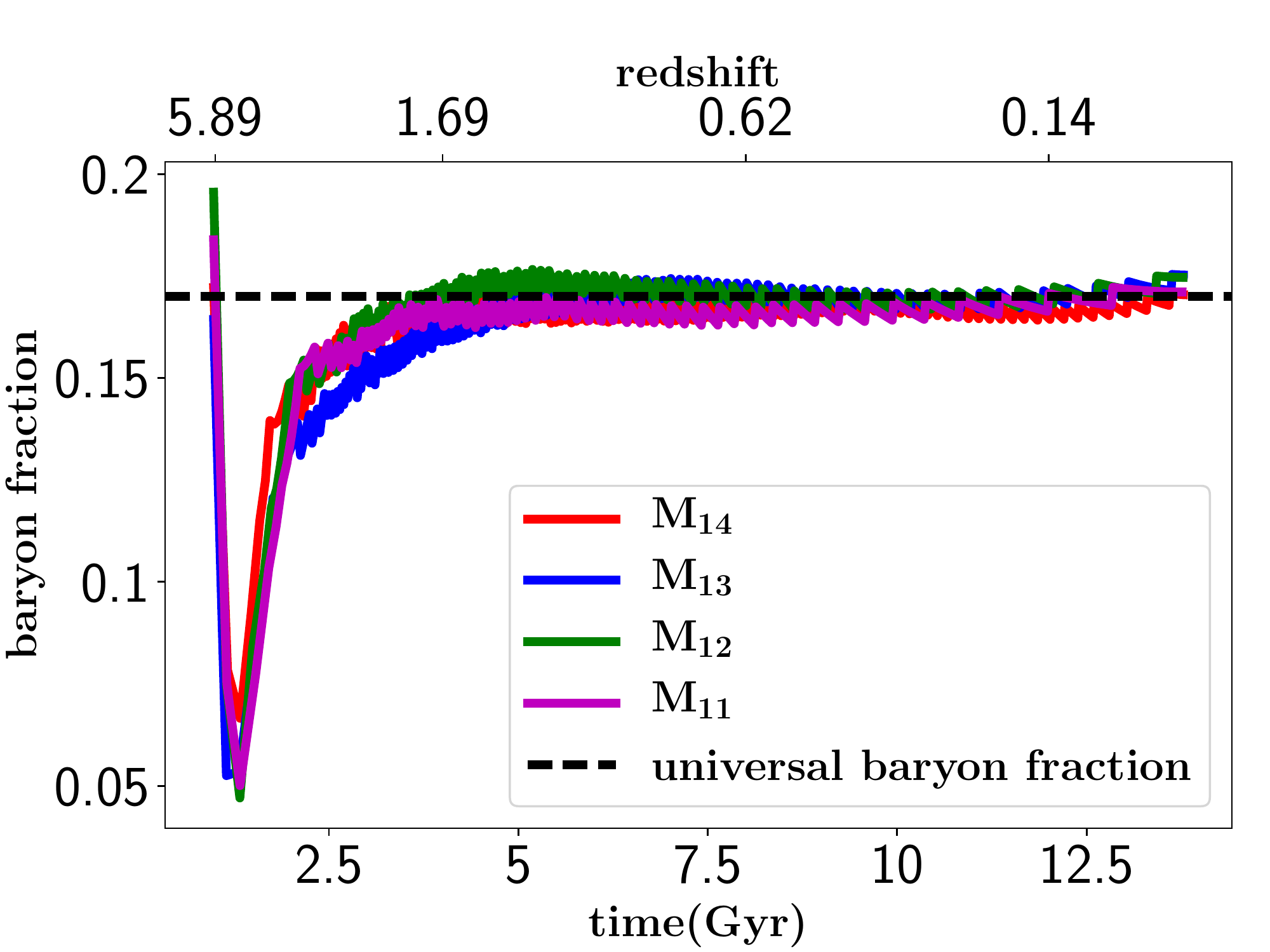}
 \caption{The gas fractions (see section \ref{sec:o1}) within $r_{200}$ at all times for non-radiative evolution. After around $3$ Gyr the deviation from universal baryon fraction ($0.17$) is less than $10\%$ and after $5$ Gyr it remains at the universal value. The initial dip in the baryon fraction reflects a transient as the gas adjusts to a self-similar profile.}\label{fig:fig04} 
 \end{figure}

Note that, $\rho_{r_{200}} = \rho_{HSE}(r_{200})$ by Eq. \ref{eq:match} and $\rho_{2r_{200}} = \rho_{O}(2r_{200})$ for a given halo. The motivation for a power-law density in the outskirts is primarily the simplicity of the form and are obtained by trial and error. In order to fix $\alpha$ for a given halo, we carry out a large number of non-radiative test runs for a range of $\alpha$. Such runs are computationally inexpensive and hence very fast. This way we find the $\alpha$ for which baryon fraction follows the universal value ($\approx 0.17$) most of the time. Thus we obtain Figure \ref{fig:fig04}. 

As is clear from the values of $\alpha$ in the Table \ref{table:haloruns}, for $M_{14}$ the slope of outer density profile is maximum ($\alpha=0$). For smaller halos, the values of $\alpha$ increases slightly, making outer densities falling very slowly at very large radii. This is understandable because massive halos accrete faster than lower mass halos (Figure \ref{fig:fig01}). We also scan a range of values for the most appropriate extreme outer trajectory ($> 2r_{200}$) initially, to make sure that by current time there are enough number of shells to fall around the virial radius. The gas that we distribute outside the halo, solely serves as a reservoir from which the halo can accrete slowly with time. This outer density profile is not observationally constrained. 

\subsubsection{Number of shells}

For all the four halos we tested the non-radiative runs with different number of shells. We set up the representative runs shown in this paper with the number of shells stated in Table \ref{table:haloruns}, such that the gas fraction (as discussed in section \ref{sec:o1}) remains approximately unchanged with the increasing number of shells and the number of shells is also not so high that the computation becomes expensive. 

\subsection{Evolution of gas}
We solve the continuity, momentum and energy equations for concentric gas shells and use the ideal gas equation of state to relate pressure, density and temperature at each time-step. 
We use the numerical scheme implemented by \cite{thoul_weinberg1_94} (section 2.3), which is a standard, second-order accurate (both in space and time), Lagrangian finite-difference scheme, to solve the following hydrodynamic equations for Lagrangian shells:  
\ba
\label{eq:eq2}
dm &=& 4\pi r^2\rho dr ,\\
\label{eq:eq3}
\frac{dv}{dt} = -4\pi r^2 \frac{dp}{dm} &-& g(r,t) ,\\
\label{eq:eq4}
\frac{du}{dt} = \frac{p}{\rho^2} \frac{d\rho}{dt} &+& \frac{\Gamma_{\rm h} - \Lambda_{\rm c}}{\rho} ,\\
\label{eq:eq5}
p &=& (\gamma -1)\rho u,
\ea
where $p$, $\rho$, $u$ have usual meanings as described in section \ref{sec:ini}, $v$ denotes velocity of the shells, $dm$ is the mass in each shell, $\Gamma_{\rm h}$ is the energy injection rate density into the gas from a central source of heating (discussed in section \ref{sec:cool}), $\Lambda_{\rm c} = n_i n_e \Lambda(n,T)$ is the radiative cooling rate and $\Lambda(n,T)$ is the cooling function, the form of which we will specify in section \ref{sec:cool}. 
For non-radiative runs in section \ref{sec:ad}, the cooling and heating terms in eq. \ref{eq:eq4} are not included. For runs with radiative cooling (and heating), these terms are treated separately by the process of subcycling which is referred to in the next section, section \ref{sec:cool}.The code time-step at each time is determined by finding the minimum of the dynamical time, Courant time and shell-crossing time calculated for all the shells (as described in section 2.4 of \citealt{thoul_weinberg1_94}).

\subsubsection{Radiative cooling}

\label{sec:cool}

\begin{figure*}
\centering
 \includegraphics[width=1.05\textwidth]{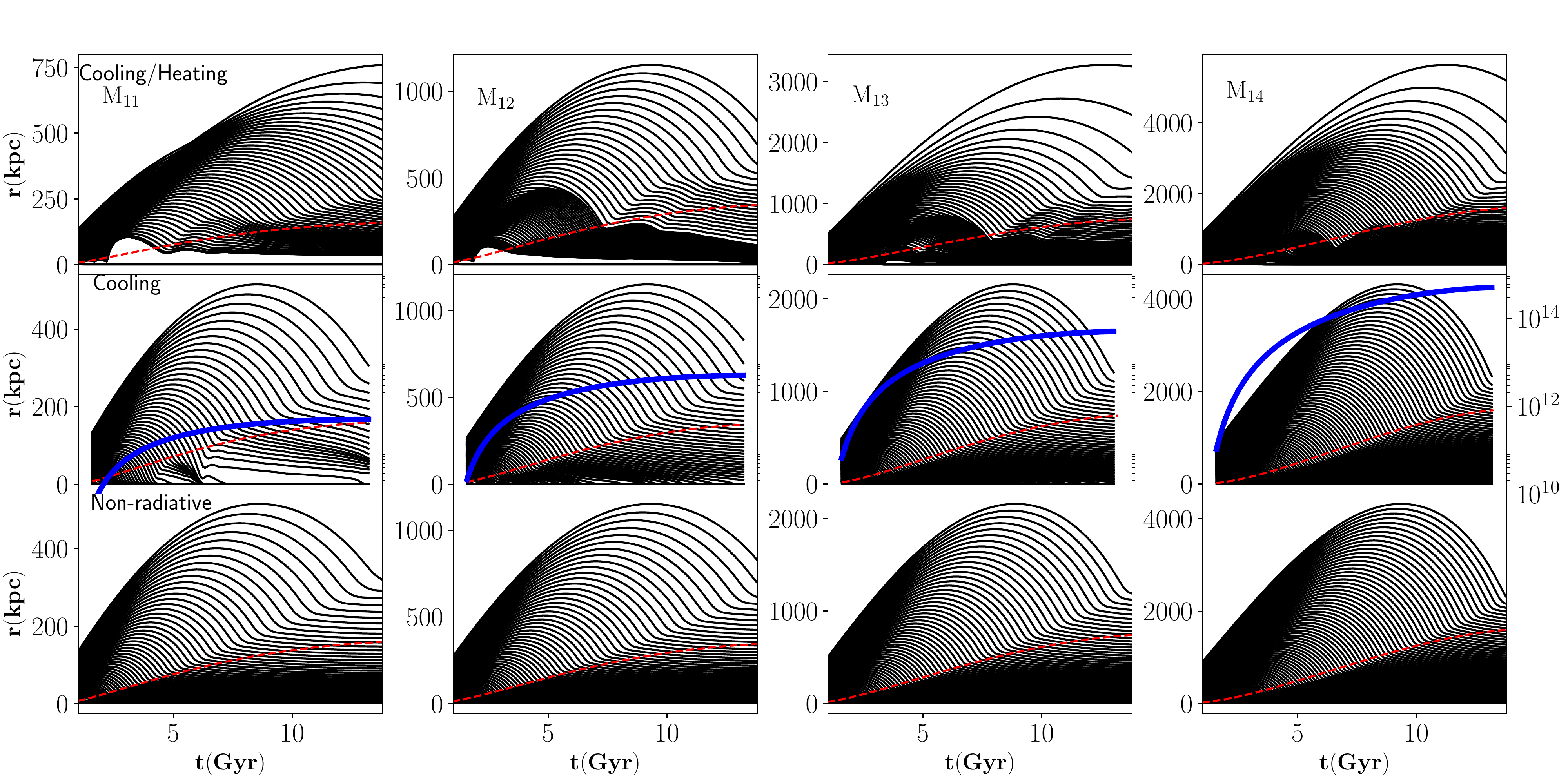}
 \caption{The trajectories of gas shells as a function of time (for clarity only a few trajectories are plotted for each halo) for all the runs. Each column shows different evolution (non-radiative, only cooling, cooling+heating) for a single halo. Lower panels show the non-radiative case, middle panels show the cases with only radiative cooling and the upper panels show the cases with cooling and heating. The dashed red lines show the $r_{200}$ that we can evaluate for the chosen mass accretion history. In the middle panels the halo mass (in $M_{\odot}$) from the mass accretion history is shown in solid blue line as a function of time (the ticks on the rightmost panel corresponding to $M_{14}$ gives the scale).}\label{fig:fig02} 
 \end{figure*}
 
 \begin{figure*}
\centering
 \includegraphics[width=0.9\textwidth]{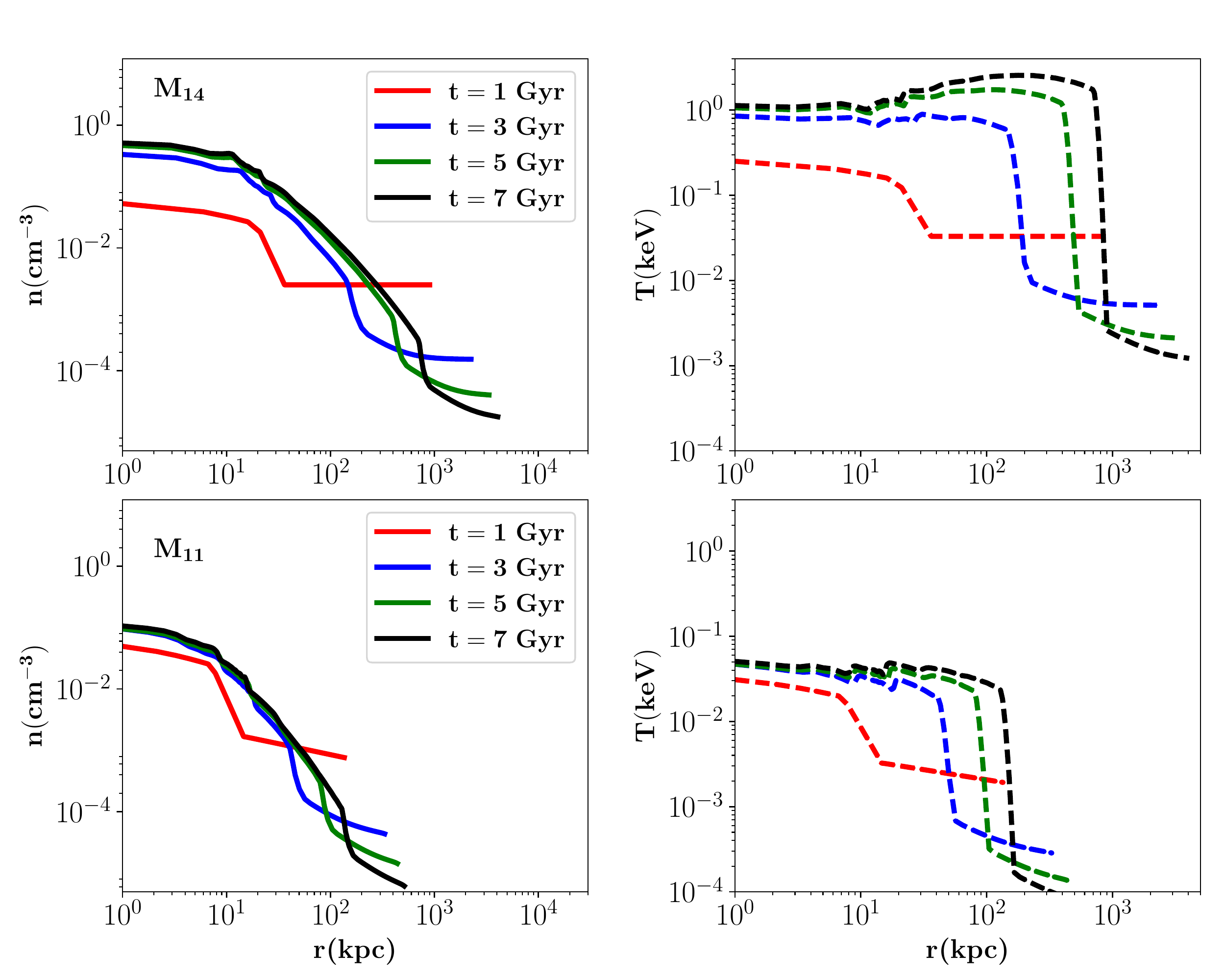}
 \caption{The evolution of number density (left) and temperature (right) for runs $M_{14}$ and $M_{11}$. The red lines show the initial conditions as described in section \ref{sec:ini}. The blue, green and black lines show the profiles at later times. The IGM gas outside the virial radius has very low temperatures at later times due to adiabatic expansion and the region close to the virial radius has the maximum temperature.}\label{fig:fig03} 
 \end{figure*}
 
\begin{figure}
\centering
 \includegraphics[width=0.5\textwidth]{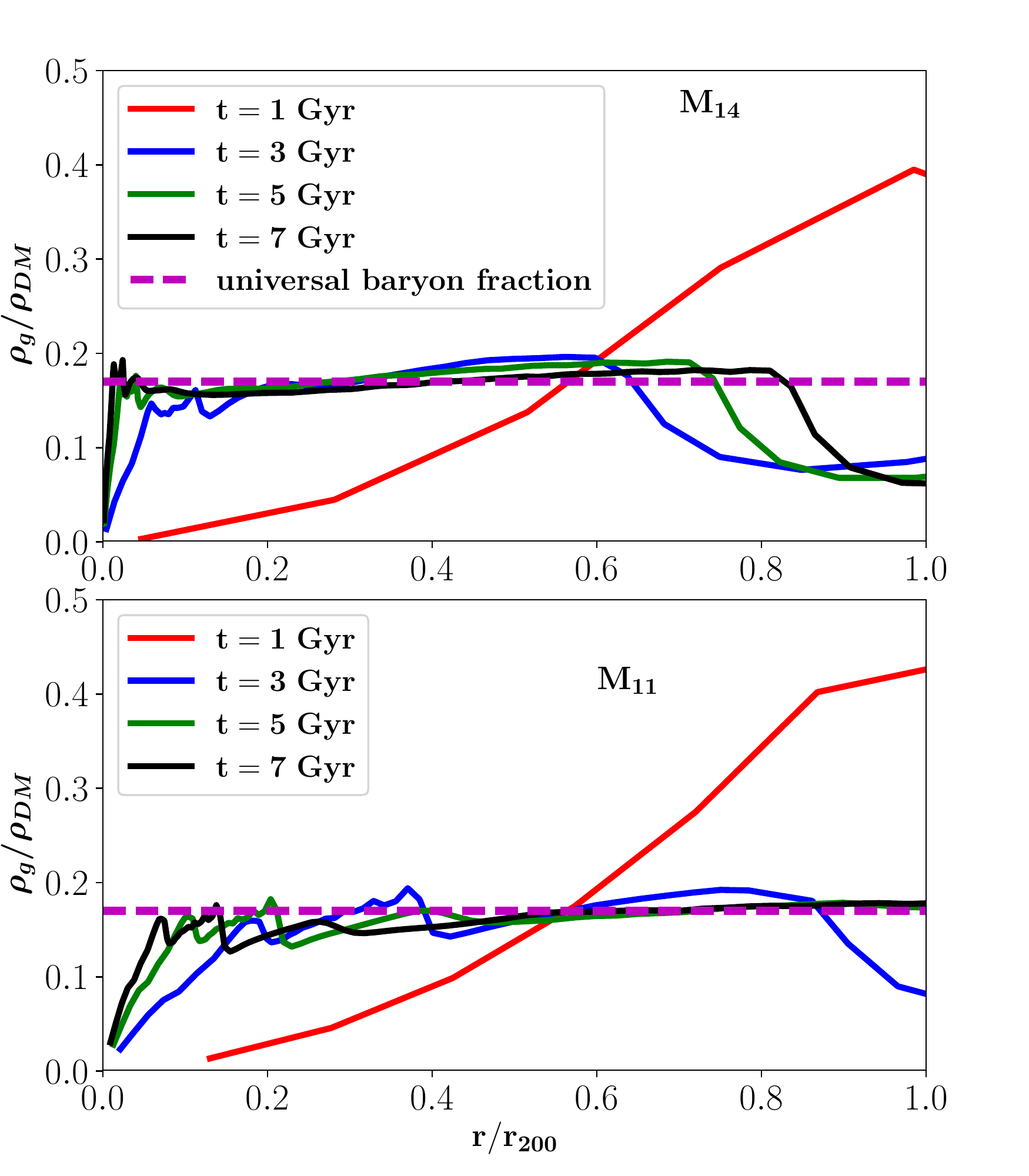}
 \caption{The evolution of the ratio of gas density to dark matter density (NFW) with radius at different times for $M_{14}$ (upper panel) and $M_{11}$ (lower panel) under non-radiative conditions. The red line shows the initial hydrostatic equilibrium and this clearly does not follow the dark matter although the total gas to dark matter mass ratio within $r_{200}$ is around the universal value. Subsequently, gas follows the dark matter within the halo at almost all radii. }\label{fig:figgasdmrat} 
 \end{figure}
 
 \begin{figure}
\centering
 \includegraphics[width=0.5\textwidth]{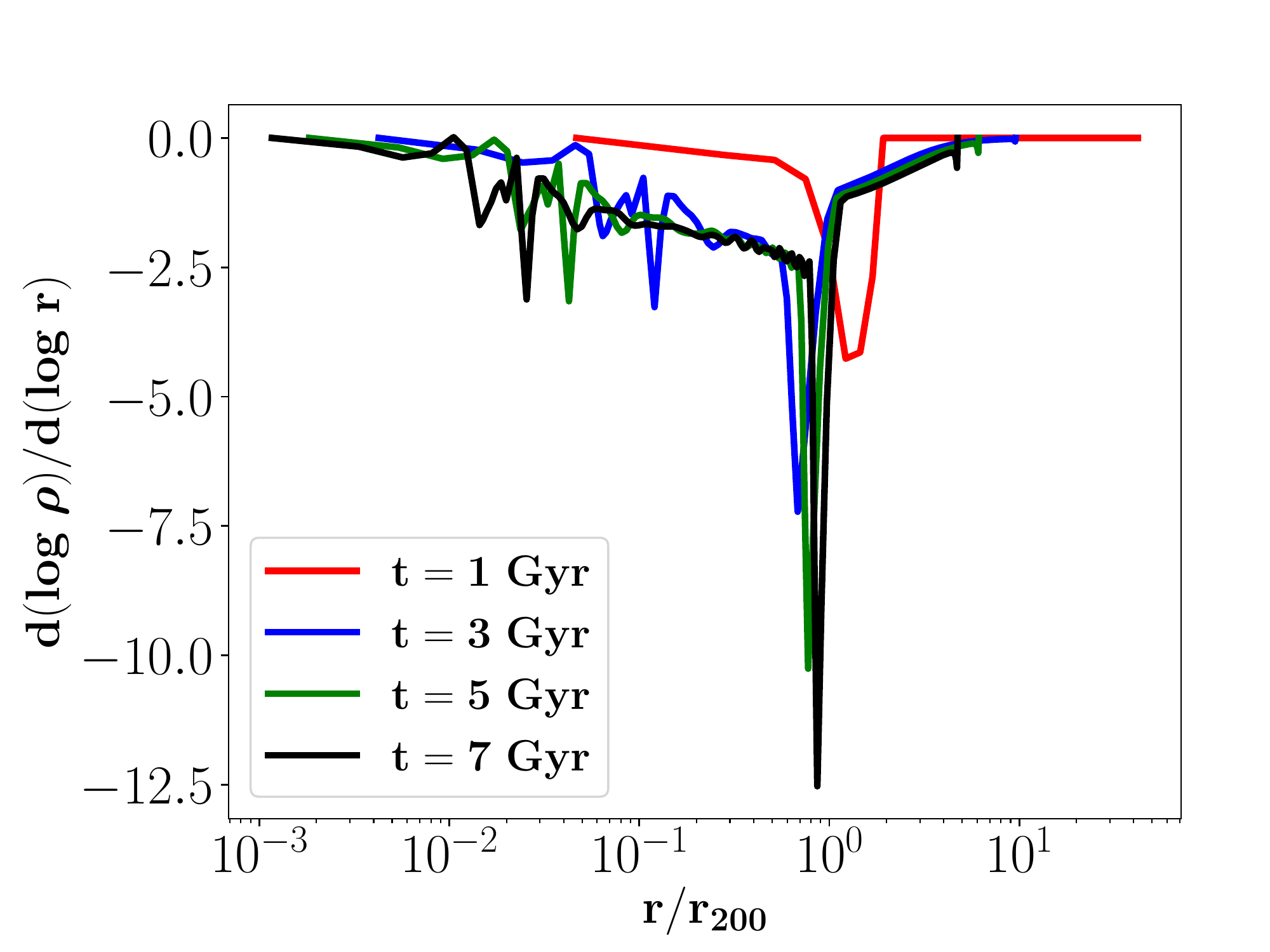}
 \caption{The density power-law index as a function of normalized radius for the biggest halo in our model ($M_{14}$). The red line corresponds to the initial condition. The halo evolves almost self-similarly at late times irrespective of the initial condition.  The large negative power law index corresponds to the virial shock. }\label{fig:figgrad} 
 \end{figure}

For runs with radiative cooling, we use an analytic fit to the cooling function corresponding to zero metallicity (see blue dotted line in Figure 1 and Eq. 6 from \citealt{Wang2014}). We cool the gas using subcycling at every hydrodynamic step. This means that at each hydrodynamic time-step, we run a loop for $N$ number of times where $N = dt/dt_{\rm cool,min}$ and $dt_{\rm cool,min}$ is the minimum of the cooling times of all the shells and update the internal energy ($u$) with the last term in Eqn \ref{eq:eq4}. This way the main time-step is not too short. We combine the cooling and heating terms in which the latter is simply zero for pure radiative cooling runs. We use a first-order explicit (semi-implicit) method to update the specific energy of the gas shells in each cooling step if heating (cooling) dominates (\citealt{sharma10}). Cooling is turned on only inside $r_{200}$ at each time, although the shock radius may be larger than $r_{200}$ at late times. The gas outside the virial shock has quite low temperature in this model and would not be affected by cooling in any case. 

\subsubsection{Heating prescription}
\label{sec: heat}
For runs with feedback heating, we use an idealized Bondi-Hoyle-Lyttleton (\citealt{bondihoyle1944}, \citealt{1939hoylelyttleton}, \citealt{1952MNRAS_bondi}) accretion rate to compute the heating rate. This is an idealized treatment of heating and Bondi accretion rate is not necessarily the best estimate for black hole accretion rate (\citealt{2013MNRAS_gaspari}, \citealt{2017MNRAS_prasad}). However, what is more relevant for our purposes in this model, is the total energy injected by feedback (as long as most of it couples to the CGM) instead of the specific feedback prescription. The following equations and parameters are used to implement heating.
\ba
\nonumber
\dot{ M}_{\rm BHL} &=& \frac{4\pi G^2 {M}^2_{\rm BH} \rho}{(c_s^2 + v^2)^{\frac{3}{2}}},\\
\nonumber
\dot{M}_{\rm Edd} &=& \frac{4\pi G M_{\rm BH}m_p}{\epsilon_r \sigma_T c},\\
\dot{E}_{\rm feed} &=& \epsilon_f {\rm min}(\dot{M}_{\rm BHL}, \dot{M}_{\rm Edd}) c^2,
\ea

where $\epsilon_r = 0.1$ is the radiative efficiency and $\epsilon_{\rm f} $ is the feedback efficiency which we adjust to obtain the baryonic properties of the halos. For calculating BHL rate we use the density ($\rho$) and velocity ($v$) of the first active (i.e. non-frozen; explained in the next section) shell and trajectory respectively. The first active shell is just outside $r_{\rm min}$. Note that the feedback power is limited by the Eddington limit. The total energy that is injected within an injection radius $R_f$ (which we take as a parameter; third last column in Table \ref{table:haloruns} shows the values we use for $R_f$) in time $dt$ is $\dot{E}_{\rm feed} dt$. The energy injection rate per unit volume is chosen to be constant, $\dot{E}_{\rm feed}/\frac{4}{3} \pi {R_f}^3$, in each shell within $R_f$. Also, for different halos we have used different values of injection radius ($R_f$) and efficiency ($\epsilon_f$) to match the observed stellar mass versus halo mass relation (values given in Table \ref{table:haloruns}) as discussed in the results (section \ref{sec:feedback}).


\subsubsection{Freezing of shells in radiative runs and floor temperatures in runs with feedback}
\label{sec:freez}
In the runs with only radiative cooling, shells come very close in the central region and the hydrodynamic time-step ($ dt \propto dr$) reduces to very small values for all the runs with cooling (including the ones with feedback). In order to get rid of this difficulty, whenever the first active shell has temperature ($T < T_{\rm thresh}$) where $T_{\rm thresh} = 2\times 10^4$ K (for $M_{14}$ and $M_{13}$) and $T_{\rm thresh} = 1.8\times 10^4$ K (for $M_{12}$ and $M_{11}$) and it reaches $<1.05r_{\rm min}~(1.05~{\rm kpc})$, we put them at $r_{\rm min} = 1$ kpc and set their velocity to zero and no longer evolve them. This way the exact value of the smallest hydrodynamic time-step is typically $3-4$ orders of magnitude less than $t_{z=6}$. For lower mass halos, the entire halo is cold at $z=6$ and there is no pressure support essentially at the beginning of the run. The shells fall at free-fall rate. With this freezing condition, we are further aiding the falling of shells. Hence it is important to keep $T_{\rm thresh}$ low enough so that a large fraction of halo gas is not artificially frozen and high enough so that we can reduce the run-time significantly. So we reduce $T_{\rm thresh}$ slightly in small halos, which have much lower virial temperatures at $z=6$, to be in the right regime. Similar techniques have been used in 1D models to avoid very short time-steps (\citealt{thoul_weinberg1_94}, \citealt{1997astroph_mirowhite}).

We have tested $M_{11}$ with different $T_{\rm thresh}$ to gauge its role. Reducing freezing temperature below what we use, will enhance the run time and will change the total amount of cold gas ($\approx 10^4~ {\rm K}$) by $\sim10\%$ and even less for massive halos. It is worth noting that in our runs with both cooling and feedback heating, which is the most realistic scenerio, very few shells (typically $< 10$) are frozen and the effect of freezing is completely negligible. Additionally, in order to verify our freezing conditions quantitatively, we compare the time-averaged hot gas content in the SAMs and the galaxy cluster in our model ($M_{14}$) and get a close match (discussed in section \ref{sec:radprof}). 

In the runs with feedback, the low-density bubble/cavity in the central region of the halo compresses shells around it and this reduces the hydrodynamic time-step significantly. These shells may cool out and come even closer which is slightly controlled by keeping a floor temperature. We set a floor cooling temperature for different halos, below which we do not cool the shells (second last column in Table \ref{table:haloruns}), only for the runs that include feedback heating.  

\section{Results} 
\label{sec:res}
A summary of all the different initial parameters tested for different halos is given in Table \ref{table:haloruns}. 
\begin{table*}
 \caption{A summary of the halo models and different initial conditions, cooling and heating parameters}
{\centering
\begin{tabular}{| c c c c c  | c c c c c|}
\hline\hline
\multicolumn{5}{| c |}{Non-radiative parameters}&\multicolumn{5}{c |}{Cooling+heating parameters}\\
\hline
Model & Current halo mass & Number of shells& Initial outer &$\alpha$ &Shells frozen&$\epsilon_{\rm f}$&$R_f$&Cooling&Seed black\\
&($M_{200}(z=0)$)&&shell radius&&at (K)&&(kpc)&stopped&hole mass\\
&($M_{\odot}$)&&(kpc)&&in runs&&&at (K) in runs&$(M_{\odot})$\\
&&&&&with cooling&&&with heating\\
\hline
$M_{14}$ & $5 \times 10^{14}$ &$180$&$901.0$& $0.0$&$2\times10^4$&$0.3$&$27$&$5\times10^4$&$10^8$\\
$M_{13}$ & $5 \times 10^{13} $&$150$&$501.0$& $0.03$&$2\times10^4$&$0.08$&$18$&$4\times10^4$&$10^8$\\
$M_{12}$ & $5 \times 10^{12}$&$135$&$271.0$& $0.24$&$1.8\times10^4$&$0.03$&$18$&$3\times10^4$&$10^5$\\
$M_{11}$ & $5 \times 10^{11}$ &$120$&$136.0$& $0.36$&$1.8\times10^4$&$0.005$&$12$&$2\times10^4$&$10^5$\\
\hline\hline
\end{tabular}
}
\\ \textbf{Notes:} All the models are tested excluding/including radiative cooling and heating. The inner most radius ($r_{\rm min}$) is fixed at $1.0$ kpc for all runs; we calculate the Bondi-Hoyle-Lyttleton accretion rate based on the active shell just outside this radius. The initial 
outer shell radius is selected in the non-radiative runs such that almost all the shells join the halo by the end of the evolution.
 \label{table:haloruns}
\end{table*}

\subsection{Non-radiative runs}

\label{sec:ad}
The non-radiative runs reflect the effect of the growth of dark matter halos on the inflow of gas, and an almost self-similar evolution of the gas density profiles. A temperature gradient develops over time within the halo, with a characteristic non-isothermal profile.

The lower panel of Figure \ref{fig:fig02} shows the non-radiative evolution of shell trajectories of all the halos we consider. The shock radius moves out with time and the gas shells get shock-heated and settle in almost parallel layers around half the turn-around radius, after they join the halo, indicating that the shells are close to hydrostatic equilibrium. At later times the shock radii of all the halos are slightly greater than the corresponding $r_{200}(t)$. 

Figure \ref{fig:fig03} shows the time evolution of the number densities and temperatures of two representative small and large halos for different times. The most massive halo, corresponding to a cluster, contains hot gas the peak temperature of which goes upto a few keV. The smallest halo show a lower overall temperature which is consistent with a self-similar evolution. The virial shock and the corresponding jump in the halo temperature are distinctly seen in all our halos. 

The temperature in $M_{14}$ and $M_{11}$ declines inward and becomes flatter, respectively. The gas entropy profile in a halo results from its assembly history and plays a key role in setting the temperature of the gas (e.g see Eq. 1 and Figure 3 in \citealt{2013MNRAS_mccourt}). Non-radiative cosmological simulations of galaxy clusters show that the temperature becomes flat toward the center and does not quite fall as in our M14 run (e.g. Figure 1 in \citealt{2007ApJ_nagai}). A comparison of non-radiative SPH and grid-based cluster simulations show that the temperature profile in the former (older SPH codes typically suppress mixing) does not match with the latter (Figure 1 in \citealt{2003MNRAS_ascasibar}). This highlights the importance of mixing in a realistic cosmological assembly of galaxy clusters. Our 1D simulations do not explicitly account for fluid mixing and therefore have temperature profiles that just reflect the mass accretion history, and do not exactly match the cosmological simulations. Our focus is on cooling and heating and this feature of non-radiative 1-D simulations will be profoundly affected by these processes. 



Figure \ref{fig:figgasdmrat} shows the ratio of gas density to dark matter density of $M_{14}$ and $M_{11}$ at different times. The radial density profile of gas in 
hydrostatic equilibrium that we impose initially at $z=6$ is shown in red. At later times, however, inside the halo (within $r_{200}$), the gas begins to follow the dark matter gradually (except at small radii where the gas profile forms a small core). This shows that the halo is slowly accreting from the gas reservoir that we have initially outside $r_{200}$. Note that we do not impose any condition on the gas fraction outside the halo in this model. In order to analyse the baryon fraction far away from the halo, a more accurate dark matter density profile should be used (\citealt{diemer_kravt2014}). Figure \ref{fig:figgrad} shows the density gradient as a function of radius in which the approximate self-similar evolution is also quite evident. The important point to note from these two figures is that the evolution at later times is not dependent on the initial distribution of gas within the halo and the redistribution of gas from the outer region happens over time. 



Now we move to more realistic runs, starting from runs with cooling but no heating, and then to runs with both cooling and feedback heating.

\begin{figure}
\centering
 \includegraphics[width=0.5\textwidth]{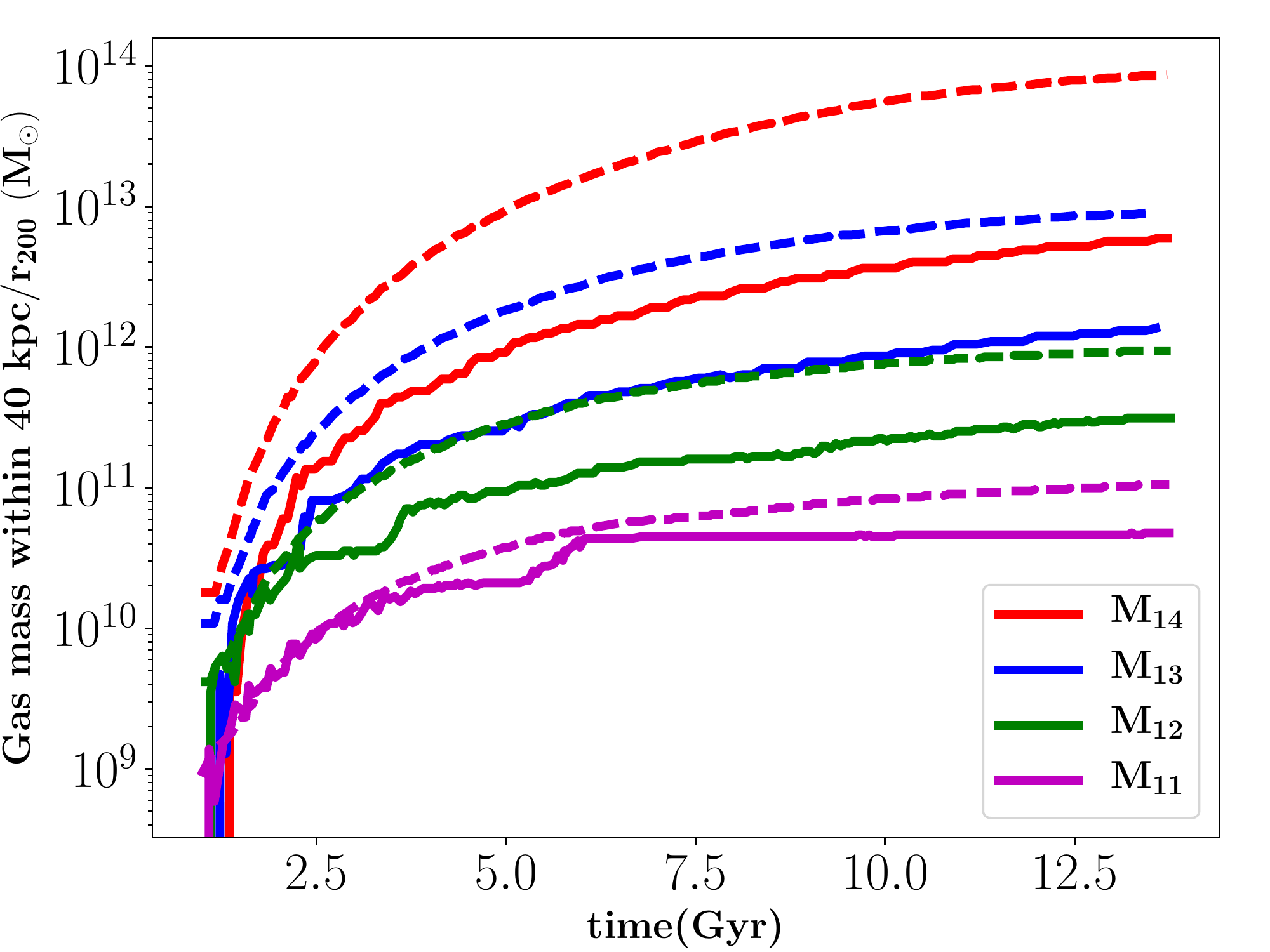}
 \caption{Mass of cold gas within $40~{\rm kpc}$ (solid lines) and total gas mass within $r_{200}$ (dashed lines) as functions of time for the four halos with only radiative cooling. The smallest halo is almost entirely cooling out in the absence of any central heating source. In contrast, the bigger halos retain relatively more hot gas.}\label{fig:fig07} 
 \end{figure}

 \subsection{Runs with only cooling}
 The middle row of Figure \ref{fig:fig02} shows shell trajectories of all the halos with radiative cooling as described in section \ref{sec:cool}.
 In the smallest halo, which has a current mass of $5 \times 10^{11}~M_{\odot}$, the virial shock does not form initially and the gas cools out and 
 falls directly to the center.
 This is the cold mode accretion, known to be prominent in halos smaller than approximately $10^{11.5}~M_{\odot}$. \cite{2003MNRAS_birnboimdekel}
 shows that the critical mass below which cold mode accretion dominates is around $3 \times10^{11}~M_{\odot}$, irrespective of the redshift. We see that in $M_{11}$,
 by the time the shock becomes stable (around $6-7~{\rm Gyrs}$), the mass of the halo is close to $10^{11}~M_{\odot}$, which is slightly less than the critical mass for the stability of virial shock. For bigger halos, the virial shock is stable at all times and all the halo masses are $\gtrsim10^{11}~M_{\odot}$ almost from the beginning (see the blue lines in the panels of the middle row of \ref{fig:fig02}). Note, however, that the inner shells are cooling and piling up in the center. Consequently the central density is very high.
   
 For all our runs with only cooling, we define gas with temperature is $\lesssim 2\times10^4~{\rm K}$ (including ``frozen" shells) to be cold. In Figure \ref{fig:fig07} we compare the total cold gas mass within $40~{\rm kpc}$(solid line) and the total gas mass within $r_{200}$ (dashed line) in all 
 the halos. In the biggest halo there is a huge reservoir of hot gas and only the gas in the central region of the halo cools out and falls to the center.
 In $M_{11}$ a large amount of gas cools out easily as the virial temperature is lower than that in the bigger halos and cooling is more efficient. This is consistent with galaxy formation models.
 
 \subsection{Runs with cooling \& feedback heating}
 \label{sec:feedback}
 In these runs we inject energy as feedback into our halos to prevent cooling flows. Feedback is triggered by the accumulation of gas close to the inner radius, $r_{\rm min}$. 
We use an idealized Bondi-Hoyle-Lyttleton prescription (section \ref{sec: heat}) for the accretion of gas. The seed black hole masses are given in the last column of Table \ref{table:haloruns}. The suppression of gas cooling depends on the total feedback energy dumped in the core (and its coupling with the gas), but not on the specific mode (AGN or supernovae) of feedback. From energetic requirement, however, we can estimate if supernova feedback is sufficient or if AGN feedback is necessary. 

 \subsubsection{Constraining feedback parameters from stellar mass-halo mass relation}
 \begin{figure*}
\centering
 \includegraphics[width=0.9\textwidth]{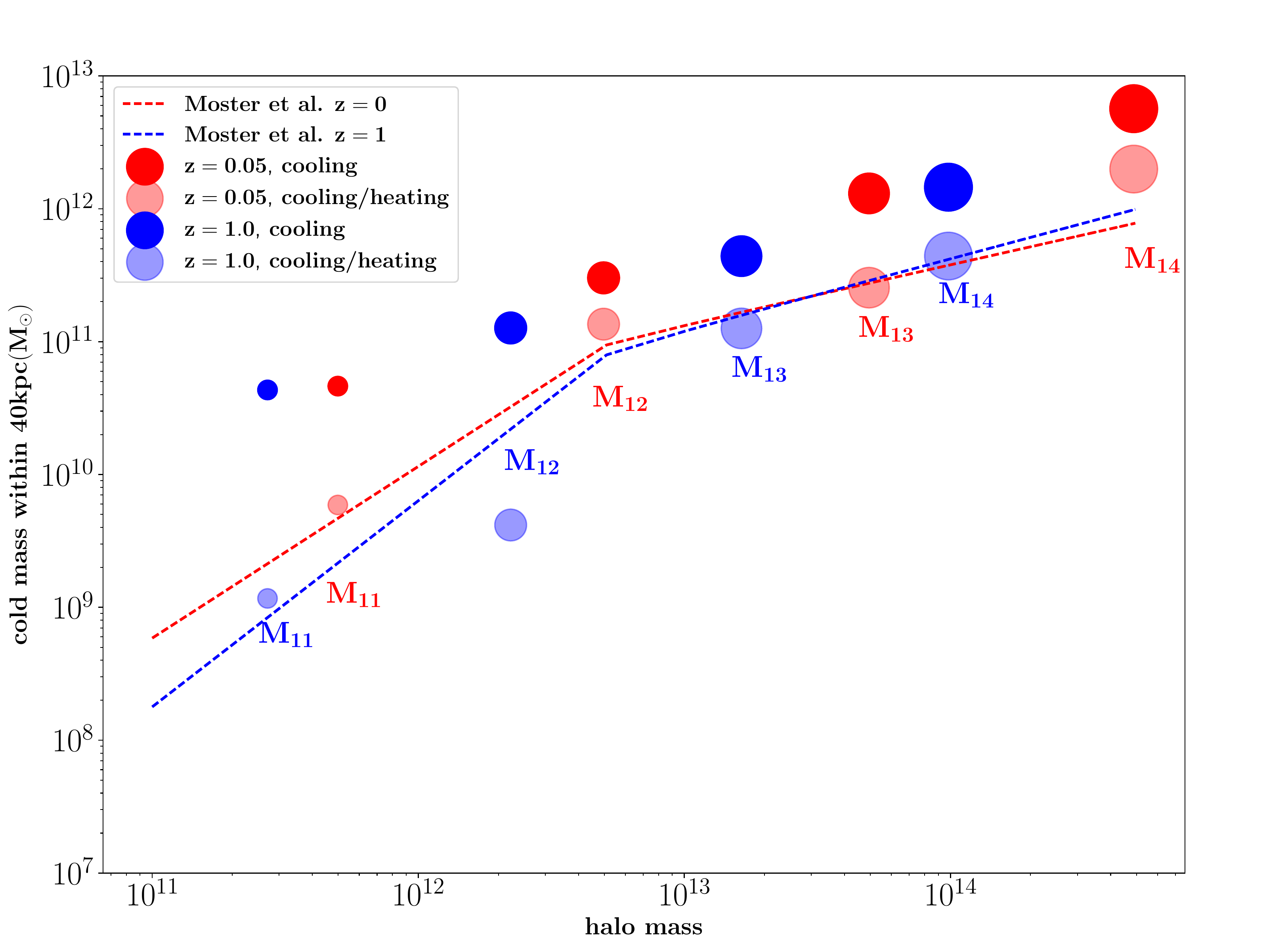}
 \caption{Cold gas mass within $40~{\rm kpc}$ versus halo mass for runs with only cooling (opaque circles) and runs with both heating and cooling (transparent circles). The runs $M_{11}$, $M_{12}$, $M_{13}$ and $M_{14}$ have current masses (in red) $5 \times 10^{11}~{\rm M}_{\odot}$, $5 \times 10^{12}~{\rm M}_{\odot}$, $5 \times 10^{13}~{\rm M}_{\odot}$ and $5 \times 10^{14}~{\rm M}_{\odot}$ respectively. The blue circles indicate the stellar/halo masses for the same halos at $z=1$.
 The dashed lines are from observations. The sizes of the circles denote the halos (for example, biggest circle denotes $M_{14}$ ). }\label{fig:fig12} 
 \end{figure*}

 \begin{figure*}
\centering
 \includegraphics[width=0.9\textwidth]{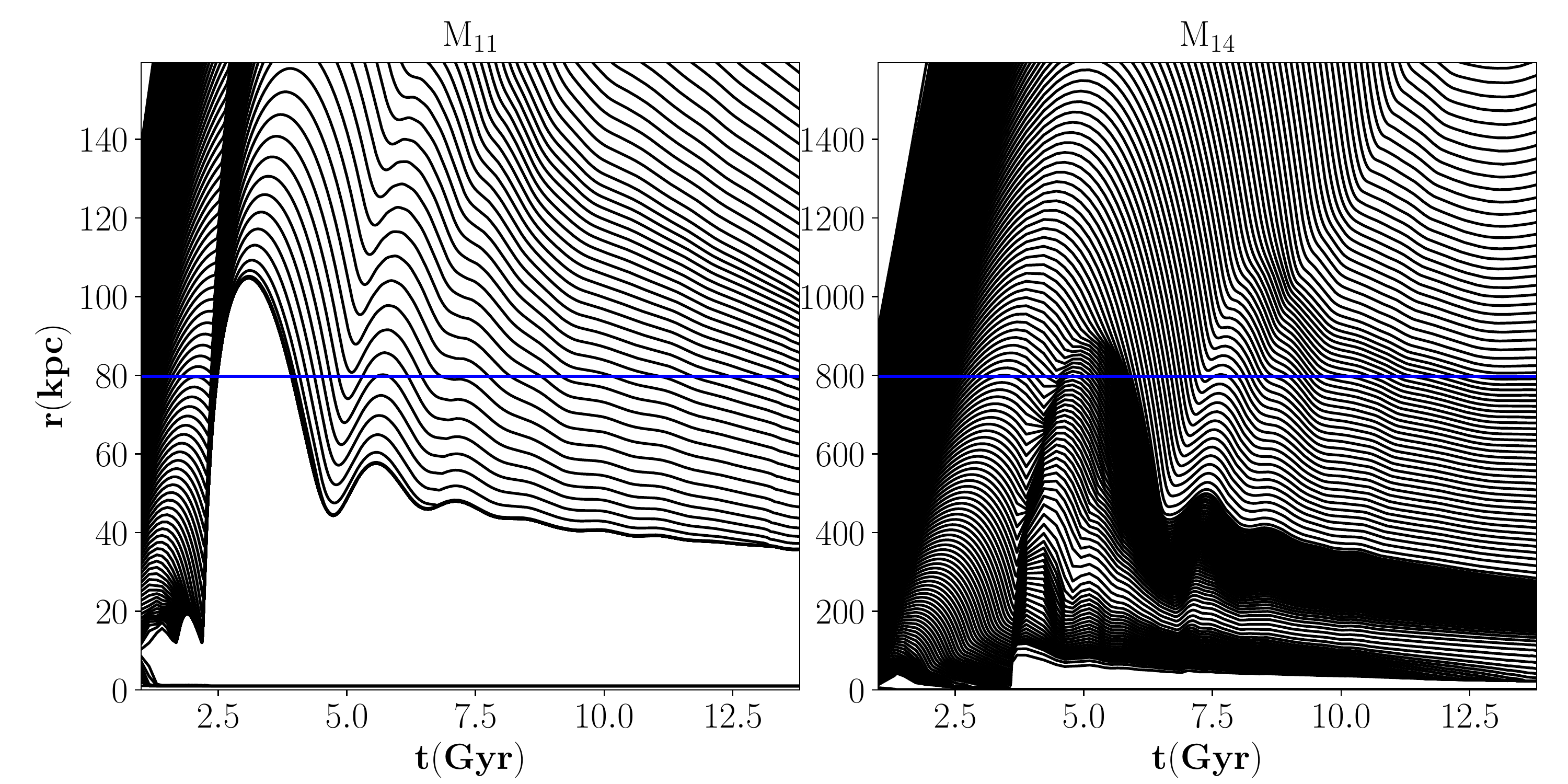}
 \caption{The trajectories of the fiducial feedback runs of $M_{11}$ and $M_{14}$ in the central region. Half of the current $r_{200}$ is marked by the blue line. The entire radial extent is upto $r_{200}(z=0)$. The smallest halo (left) shows a brief rapid accretion phase initially, which grows the black hole significantly and generates a huge feedback event. It is evacuated for most times thereafter while the biggest halo retains enough gas upto a comparable scaled radius. }\label{fig:fig09} 
 \end{figure*} 

As an estimate of the stellar mass, we use the cold gas mass within $40$ kpc. It is important to note here that only a few shells within $40~{\rm kpc}$ are frozen and most are dynamically active. This should be a good estimate of stellar mass, averaged over several dynamical times, since cooling gas ultimately forms stars. Our results do not change if we have a slightly different radius than $40~ {\rm kpc}$. Once we have an estimate of the stellar mass, we can compare our simulations to the abundance matching results. 

In Figure \ref{fig:fig12}, the underlying dashed red ($z=0$) and blue ($z=1$) lines show the average stellar mass and halo mass relation deduced from observational data and large scale simulations (\citealt{2010ApJ_moster}). On top of those we put circles from our runs, whose sizes correspond to the halo masses. For each of the four halos, we take the cold gas mass in the inner region ($40~{\rm kpc}$) as a function of time and interpolate this to find the cold mass at a specific redshift. Thus red circles and blue circles denote two different redshifts as do the dashed red and blue lines from abundance matching ($z=0$ and $z=1$ respectively). The opaque circles are derived from pure radiative cooling runs. This gives an idea of how much excess cooling happens for each halo in the absence of central heating. Accordingly, we try out a range of values for our two feedback parameters $\epsilon_f$ and $R_f$ (refer to section \ref{sec:cool} and Table \ref{table:haloruns}) so that all the circles closely follow the abundance matching results. Thus we select four sets of parameters, $\epsilon_f$ and $R_f$, for our four halos. The results for these runs are denoted by transparent red and blue circles of same sizes as those of the purely radiative cooling cases. Note that for the smallest halo, the cold gas mass has to be reduced by $2$ orders of magnitude while for clusters by only a factor less than $10$, which implies in the small isolated halos cooling is relatively more efficient. However, many small halos enter the ICM of massive halos and the gas in them get stripped by the cluster gas. In such cases, a strong feedback may not be important in such subhalos to quench star formation.

Among the two parameters that we tune, $\epsilon_f$ and $R_f$, the former is relatively more important. Note that $\epsilon_f$ is larger for massive halos because a massive halo has a deeper potential well in which it is difficult to drive out gas. This parameter determines the total energy to be deposited in the central regions which in turn would suppress the cooling. On the other hand, $R_f$ only tunes the distribution of this energy. If we vary $R_f$ slightly, the corresponding circles in Figure \ref{fig:fig09} will move above or below to some extent as we are looking at a particular instant of time. However, in the time-averaged sense, the gas evolution will be quite similar and baryon fraction cycles (discussed in the next section) will be shifted slightly across small intervals of time. $R_f$ will have a significant effect when $R_f$ is very large. The total energy gets distributed over a large volume, without forming any shock and hence the feedback does not affect the cooling flow. 

These four selected runs, for which we adjust the parameters to match the abundance matching results, are considered the fiducial runs and the corresponding trajectories are shown in the upper panel of Figure \ref{fig:fig02}. The smallest halo is completely evacuated by an early feedback event which can also be seen in the left panel of Figure \ref{fig:fig09}. There is an inward flow of gas at the very beginning which triggered the growth of the black hole according to our simple prescription. Figure \ref{fig:fig09} shows the inner $r_{200}(z=0)$ of the halo and $0.5r_{200}(z=0)$ is marked by the blue lines. The gas shells that are thrown out, are very slowly re-accreted back into the halo and there are no significant bursts after that till the current time. For $M_{12}$, there is a similar burst at the beginning, but the recovery is faster. For $M_{14}$ (and $M_{13}$) there are a few significant bursts over the entire timespan. The right panel of Figure \ref{fig:fig09} shows the inner $r_{200}(z=0)$ of the biggest halo, $M_{14}$. In this case, the halo is not much evacuated because the gravitational potential of the clusters are deeper. 

Hereafter, we investigate some of the key characteristics of our fiducial halos with radiative cooling and feedback heating.

\subsubsection{Baryon fraction evolution}
 \begin{figure}
\centering
 \includegraphics[width=0.5\textwidth]{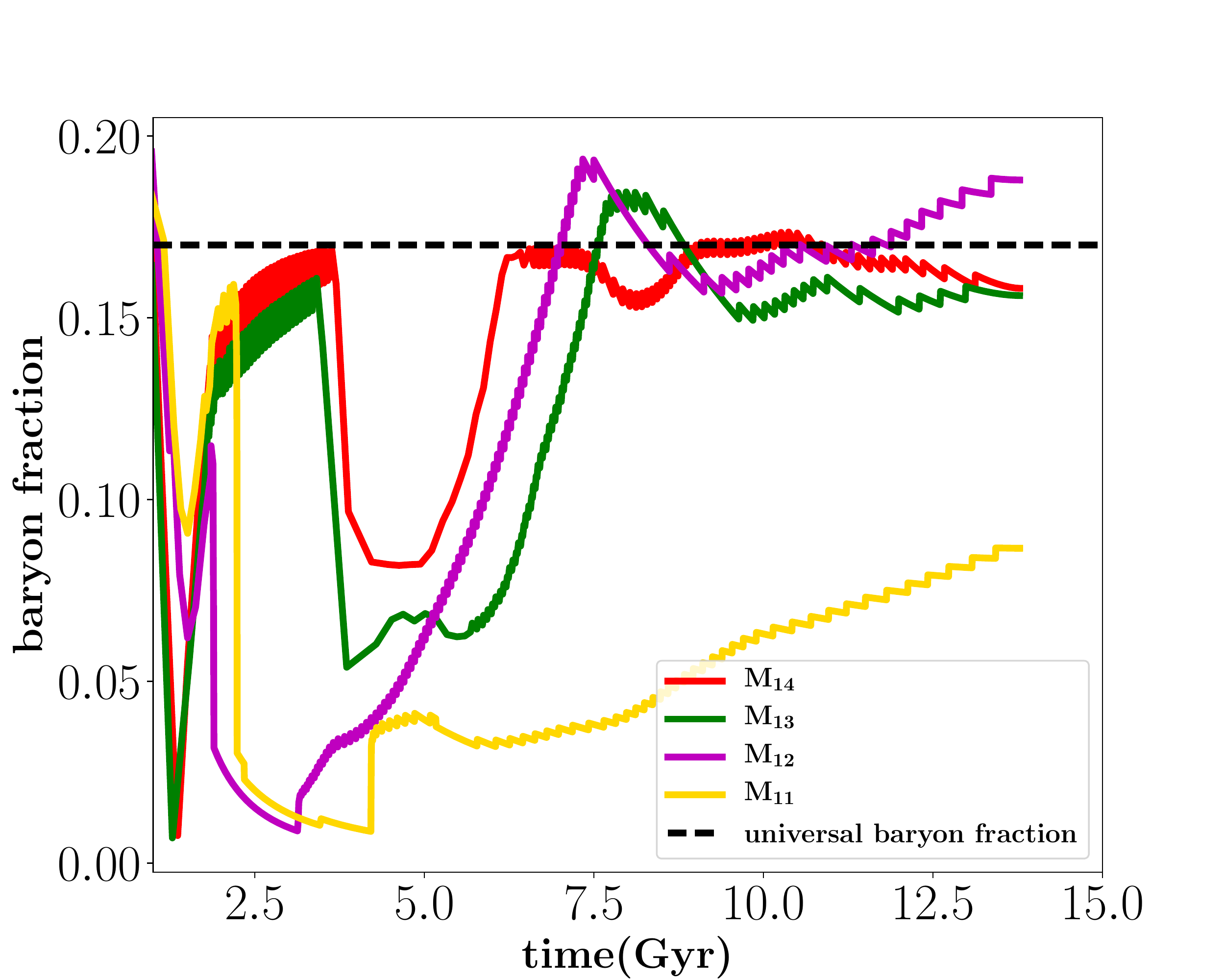}
 \caption{The gas/baryon fraction of each halo within $r_{200}$ for the fiducial cases, with radiative cooling and feedback parameters adjusted to reproduce the stellar mass versus halo mass  relation (see Figure \ref{fig:fig12}). The smallest halo is almost evacuated for the entire time and missing a large fraction of baryons due to feedback while the largest halo mostly maintains its gas content with several cycles of heating and cooling. Note that the initial dip within 1 Gyr and 2 Gyr reflects the short readjustment time the halo takes before maintaining the universal baryon fraction in non-radiative runs (see also, Figure \ref{fig:fig04}).}\label{fig:fig13} 
 \end{figure}
We define the baryon fraction as the ratio of gas mass and the dark matter mass within $r_{200}$ ($M_{200}$). Recall that, we choose the initial density profile outside the halo such that the gas fraction follows the universal value after an initial transient for non-radiative evolution (Figure \ref{fig:fig04}). Our fiducial feedback runs show deviations from the universal value and feedback may explain why baryons are often observed to be missing from smaller halos. Figure \ref{fig:fig13} shows the baryon fraction evolution for our fiducial runs. For the smallest halo, there is a large accretion event at the beginning which grows the black hole and sudden heating easily throws away most of the gas out of the halo. These gas shells take a long time to re-enter the halo and move back towards the bottom of the potential well. For most of the time the halo lacks more than $50\%$ of the baryons. For the next massive halo $M_{12}$, the gas is not thrown far out of $r_{200}$. Consequently, around $7~{\rm Gyr}$, we find the baryon fraction rising sharply. Thereafter small cycles of cooling and heating follow. In $M_{13}$ and $M_{14}$ the cycles are progressively more frequent. Particularly for $M_{14}$, only $50\%$ of baryons are missing for a brief period around $5~{\rm Gyr}$ and the halo maintains its gas content thereafter. For Milky-Way sized galaxies (our $M_{12}$ is closest to Milky Way in mass) there have been extensive discussions on whether the missing baryons are available between the virial radius and turnaround radius (\citealt{2018ApJ_bregman}, \citealt{2010ApJ_anderson}). Galaxy clusters, on the other hand, have been observed to contain the highest baryon content (\citealt{2003MNRAS_ettori}). 
 
 
\citealt{2016PhRvD_schaan} discuss the baryon content of clusters and groups by combining data from Atacama Cosmology Telescope and ``Constant Mass" CMASS galaxy sample from the Baryon Oscillation Spectroscopic survey to measure the kinetic SZ signal of the halos over the redshift range $0.4 - 0.7$. They consider the kSZ signal obtained relative to the expected kSZ signal to be a proxy for the average baryon content. In our model, $M_{14}$ and $M_{13}$ which have masses in the range of galaxy clusters and groups at current redshift respectively, show slightly more baryon content than what is predicted by them, around $z=0.5$, in the central regions. Note that these observations have several uncertainties and the proxy of the baryon fraction is only proportional to the free electron fraction $f_{\rm free}$. In our halos, the baryon fraction in the outskirts of the halo, roughly follows the universal value around that time. At smaller radii where cavity is blown out by the feedback, the average baryon fraction falls to around half the universal value. 

 \subsubsection{Radial profiles of clusters}
\label{sec:radprof}
We use our model to compare the radial profiles from $M_{14}$ and observations available from the {\it Chandra} clusters and recent Sunyaev-Zel'dovich-selected {\it Chandra} clusters (\citealt{2017ApJ_mcdonald}). 

\begin{figure*}
\centering
 \includegraphics[width=0.9\textwidth]{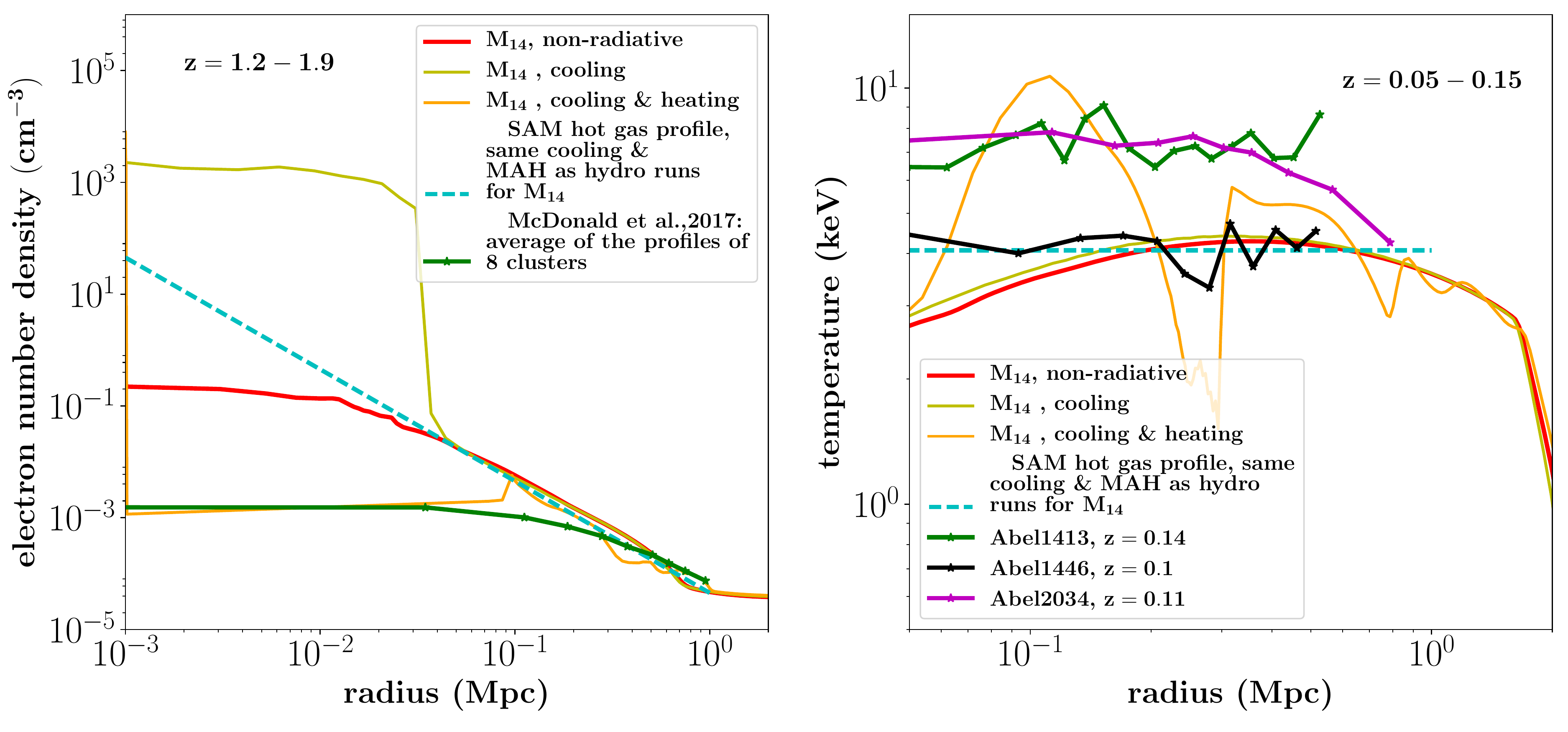}
 \caption{Left panel: Comparison of the time-averaged electron number density profiles for different evolution of the cluster run ($M_{14}$). The observation data, which is averaged for $8$ different clusters in the redshift range $1.2$ and $1.9$, is shifted along x-axis (by $0.3$ Mpc inwards to match the shock location). These are obtained from the X-ray data of the SZ selected clusters described in \citealt{2017ApJ_mcdonald}. Right panel: The time-averaged ($z=0.05-0.15$) temperature profiles of $M_{14}$ compared to the temperature profiles of the Chandra clusters at redshifts $z=0.1, 0.14, 0.11$. For both the density and temperature profiles, the dashed cyan lines show the profiles corresponding to an isothermal gas profile used in SAMs. It is useful to note that the SAM density is only for the hot gas while our runs have contribution from both hot and cold gas. Although the SAM density (subtracting the cold gas mass) is higher in the center, the mass in the hot ICM is comparable to what we have in our runs with pure radiative cooling (discussed in section \ref{sec:radprof}).}\label{fig:fig17} 
 \end{figure*}

Figure \ref{fig:fig17} (left panel) shows the time-averaged density profiles for different cases, within a redshift range $z=1.2-1.9$. The red line shows the density profile for pure non-radiative evolution. In the outskirts of the halo, this line coincides with the hot gas profile calculated from the SAM (cyan dashed line) which assumes isothermal gas with $\rho \propto r^{-2}$ in the halos. The match at large radii is expected with the same mass accretion history. However, in the central region, the SAM isothermal profile (that subtracts the gas cooling out) does not accurately trace out the observed gas distribution. The case with pure cooling flow (in yellow) shows a very large density in the central region, as expected. However, if we compute the time-averaged (between $z=1.2$ and $z=1.9$) hot gas content in SAM (using same mass accretion history and cooling prescription) and our model with pure radiative cooling (that is, time-averaged after subtracting the mass of gas below $ 2\times10^4~{\rm K}$ at each time), we see that the amount of hot gas is comparable, $6.84 \times 10^{12}~{\rm M}_{\odot}$ and $6.06 \times 10^{12}~{\rm M}_{\odot}$ respectively. The orange line shows the fiducial case with radiative cooling and feedback heating. It almost coincides with the average density profile obtained from observations of $8$ clusters (of different masses and redshifts), within $z=1.2-1.9$, except for tiny wiggles. The fiducial run also shows that some gas cools and falls to the center ( and gets frozen according to our prescription), with a peak density at $0.001~{\rm Mpc}$.

 The right panel of Figure \ref{fig:fig17} shows the comparison of the time-averaged temperature profiles for $M_{14}$ within $z=0.05-0.15$ with those of the clusters at redshifts $z=0.1, 0.14, 0.11$.  The radial profile of Abell 1446 at $z=0.1$ coincides with those of $M_{14}$ within the radial range of $0.01~{\rm Mpc}$ to $0.1~{\rm Mpc}$. However, the profile corresponding to the non-radiative run, has lower temperatures near the center. The fiducial heating run of $M_{14}$ has a temperature profile that varies around that of Abell 1446.  The SAM temperature profile falls almost on top of Abell 1446. The two clusters on the higher side of this redshift range, have around twice the temperature of Abell 1446 and $M_{14}$. The fiducial case also shows drastic temperature changes in the central core because of the shock-heated shells and rapid cooling of the compressed shells around the cavity. In the 3D extension of this model, the temperature will be shell averaged and hence will be smoother because of the presence of multiphase (both cold and hot) gas in the same shell (\citealt{prasad15}, \citealt{choudhury16}, \citealt{2017MNRAS_fielding}). The physics of the cooling blobs rather than the monolithic collapse of the entire shell, is missing in our 1D runs. The cavity temperature is as high as $10~{\rm keV}$. But our temperature profiles approximately reproduce the observations particularly in the outskirts of the halo. Hot gas profiles of density and temperature could be better compared with our future multidimensional simulations. 

 \subsubsection{Qualitative changes in the equation of state}
 
 \begin{table*}
 \caption{Parameters of the fitted lines for the EoS in $M_{14}$ at $z=0$ as $log_{10}(p) = a + \Gamma log_{10}(\rho)$}
 
 {\centering
\begin{tabular}{| c c c | c c c c |}
\hline\hline
\multicolumn{3}{| c |}{Non-radiative parameters}&\multicolumn{4}{c |}{Cooling+Heating parameters}\\
\hline
Virial shock&$a$&$\Gamma_{\rm fit}$&Virial shock&Cooling break&$a$&$\Gamma_{\rm fit}$\\
\hline
Outside& $31.43$&$1.67$&Outside&Outside&$31.72$&$1.68$\\
Inside& $13.73$&$0.93$&Inside&Outside&$18.87$&$1.11$\\
&&&Inside&Inside&$4.7$&$0.58$\\
\hline
\end{tabular}
}

\textbf{Notes:} The fitted lines are shown in black in the upper two panels of Figure \ref{fig:fig15}. The values of $a$ (the intercept) obtained from these lines are used to compute $\Gamma$ shown in the lower panels. The fitted values of $\Gamma$ ($\Gamma_{\rm fit}$) are in this table, which are the slopes of the dashed black lines.
 
\label{table:fit}
\end{table*}

 \begin{figure*}
\centering
 \includegraphics[width=0.9\textwidth]{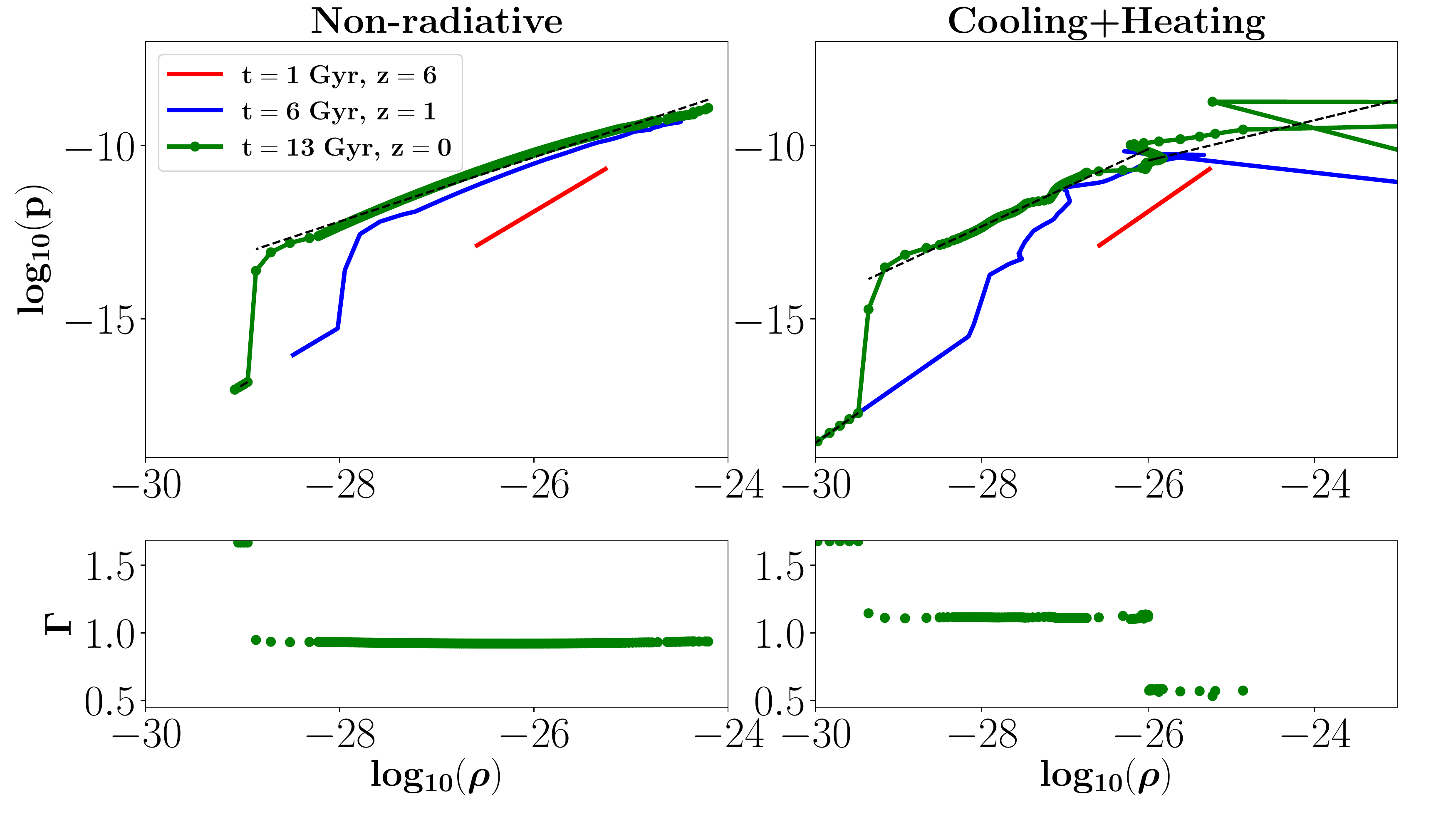}
 \caption{The pressure-density relation in cluster run ($M_{14}$) for pure non-radiative cases (left) and with cooling and heating (right) in $M_{14}$. The dashed black lines are straight lines fitted ($log_{10}(p) = a + \Gamma_{\rm fit} {log}_{10}(\rho)$) to different regimes: outside the virial shock, inside the virial shock and inside the rapid cooling zone (for the case with cooling+heating). The parameters of the fitted lines are shown in Table \ref{table:fit}. The straight lines imply that gas roughly follows  the EoS $p \propto \rho^{1.67}$ outside the virial shock, as set initially, the EoS  $p \propto \rho^{1.1}$ inside the virial shock but outside the rapid cooling zone, and approximately $p \propto \rho^{0.6}$ inside cooling zone. Lower panel shows the predicted values of $\Gamma$ at $z=0$ from our runs using the fitted parameter $a$ from the dashed lines. }\label{fig:fig15} 
 \end{figure*}

One of the useful ways to understand the pressure and density profiles for galaxy clusters is to relate these by a known equation of state of the medium. The assumption of an EoS is helpful to construct the gas profiles once the global SZ signal or the total X-ray luminosity is obtained.  

We try to see the qualitative changes in the equation of state of the gas in the ICM ($M_{14}$); specifically how $\Gamma (\equiv d\ln p/d\ln \rho)$ changes for the gas that has fallen into the dark matter halos. Figure \ref{fig:fig15} (left column) shows the profiles of the best fitted $\Gamma$ for the pure non-radiative evolution of $M_{14}$. The initial index is $5/3$, which is clearly seen by the slope of the red line (at $z=6$) and that of the lines outside the shock. As the gas gets shock-heated, the index falls to around $1.0$ in the non-radiative case and around $1.1$ in the cooling+heating case. The pressure-density relation remain invariant (fall almost along the dashed line fitted) except in the core in the cooling+heating case. Figure \ref{fig:fig15} (right column) shows qualitatively the flattening of the pressure-density relation at higher densities for evolution with cooling+heating. Radiative cooling and feedback heating alter the self-similarity in the central cores of dark matter halos.  

A simple theoretical model of ICM, that matches with hydrodynamic simulations, has found the polytropic index to be $1.15$ (\citealt{2005ApJ_ostriker}) except inside the rapid cooling zone, while X-ray observations have found the index to be around $1.2$ (\citealt{2005ApJ_solanes}). This agrees with our results. Recently \citealt{2017ApJ_flender} describe the ICM at $z=0$ by a broken power-law for pressure-density relation. The cooling break is seen around $\rho_{500{\rm c}}$, below which the index falls to around $\gtrsim 0.1$ (varying systematically with redshift) inside the core. By construction, in 1D the entire shell cools out or gets heated. Hence there is a large scatter in the pressure density relation in the core in our current model. 

\subsubsection{Growth of black holes}

 \begin{figure}
\centering
 \includegraphics[width=0.46\textwidth]{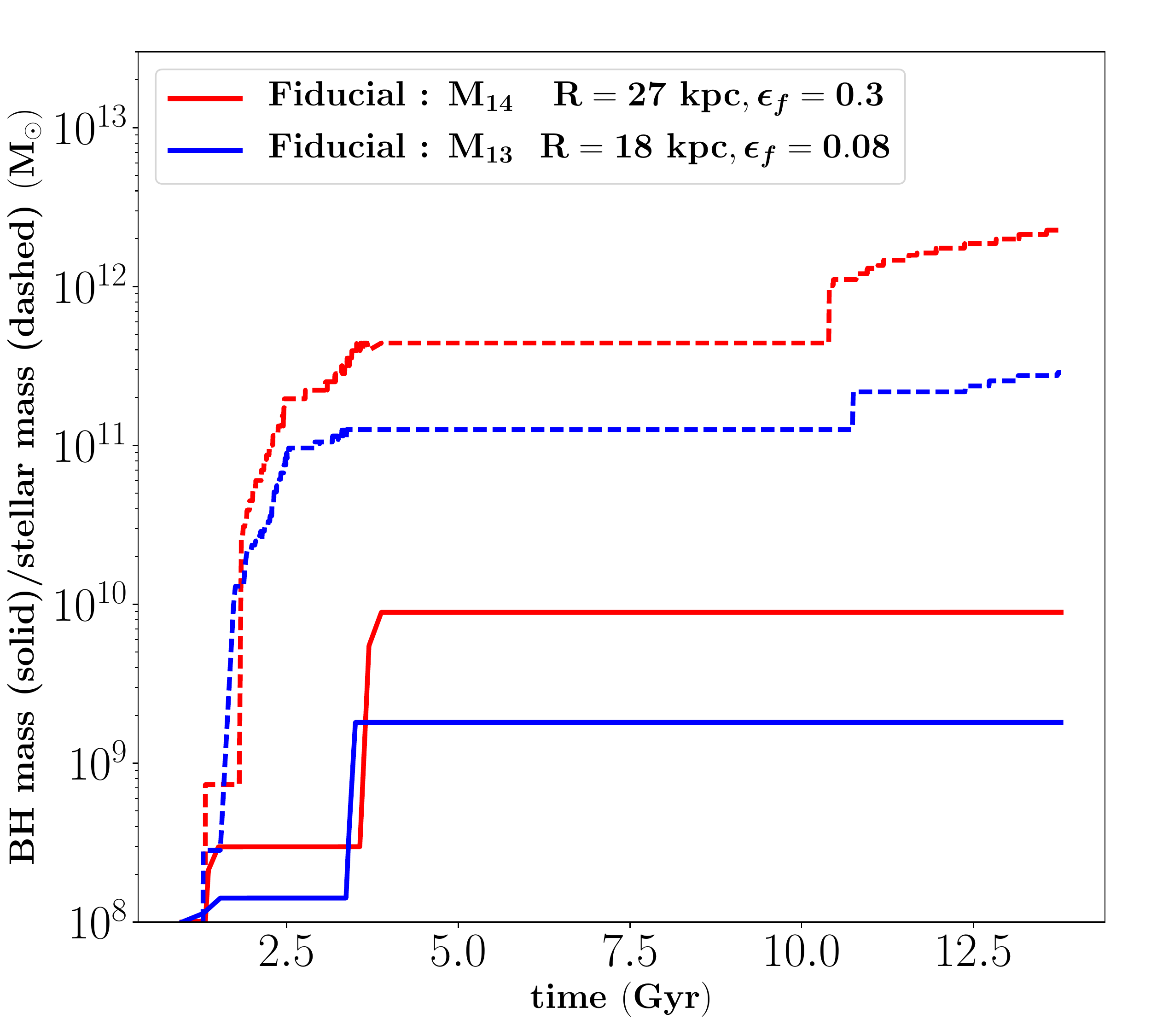}
 \caption{The black hole mass as computed from the Eddington limited mass accretion rate in our idealized spherical Bondi model (solid lines) and the cold mass within $40~{\rm kpc}$ (which is a proxy for an average stellar mass in our model; dashed lines) as functions of time. The discrete growth of black hole mass will be smoothed out in a multidimensional model with feedback inflated hot cavities in specific directions.}\label{fig:fig18} 
 \end{figure}

In our model we constrain the cold gas formation within $40~{\rm kpc}$ at $z=1$ and $z=0$ by adjusting feedback parameters. The assumption of spherical Bondi accretion (Eddington limited) 
gives a crude estimate of the accretion rate. With this accretion rate, we can also get an estimate of the approximate growth of black hole, starting from a seed. It 
must be noted that the multidimensional version of our model, with jet inflated cavities along a specific direction, will 
give a more accurate picture and a smoother growth of the black hole. Additionally, small halos (satellite galaxies), which are bound to a massive halo, will have a different evolution and growth of central black hole due to the active role of physical processes like gas stripping, tidal stripping, etc. 

In our model, the small halos $M_{11}$ and $M_{12}$ show a brief growth of black hole at around the Eddington rate, quite early in time. Thereafter, they do not have any significant growth. In \cite{2010Aj_willot}, the authors present the data for quasars around $z=6$ from Canada-France High-z Quasar Survey (CFHQS). They find that most quasars are accreting at close to the Eddington rate, which gives an exponential growth. 
Earlier works, like \citealt{2002MNRAS_yu}, emphasize that many of the massive black holes grow mostly in the bright QSO-phases. 

Figure \ref{fig:fig18} shows the black hole mass (solid lines) and the cold gas mass within $40~{\rm kpc}$ for different times in the current group and cluster scale halos of our model. This shows a similar average evolution for group and cluster. Additionally, this plot also reflects that most of the stellar mass accumulates at early times ($\lesssim$ 4 Gyr) 
and at late times separated by $\approx 5~{\rm Gyr}$ quiescence period, which is in agreement with \citealt{2018Nature_navarro}. However, for clusters with observed cavities, \citealt{2006ApJ_rafferty} argue that the black hole grows at a rate which is roughly $3$ orders of magnitude less than the star formation rate at very low redshifts. Our black holes grow at a higher average rate because of the unrealistic 1D assumption.

It is interesting to note that for the biggest halo  ($M_{14}$) which grows to cluster scale by the current time, the black hole mass goes upto $\lesssim10^{10}~M_{\odot}$, which is high compared to what is observed in clusters. This can be partly an artefact of the fact that other sources of heating like thermal conduction, falling galaxies, turbulence, dark matter subhalos, etc. are not taken into account. Moreover, \cite{2012MNRAS_Hlavacek} show the possibility of ultramassive black holes in brightest cluster galaxies with masses around a few times $10^{10}~M_{\odot}$. For halo masses of around $\approx 10^{12}~{\rm M}_{\odot}$, black hole masses greater than $\approx10^{9}~M_{\odot}$ have not been detected by current observations (\citealt{2016ApJ_lasker}). In our model, the final black hole mass for $M_{12}$ (with current mass $5\times10^{12}~{\rm M}_{\odot}$) is also $\gtrsim10^{9}~M_{\odot}$ (see Table \ref{table:fdbk}).

\subsubsection{Plausible sources of feedback}
 \label{sec:fdsrc}
 \begin{table*}
 \caption{Comparison of time-averaged power for different halos}
 
 {\centering
\begin{tabular}{| c c c c |}
\hline
Halo&Time-averaged feedback power&Time-averaged SN power&Final black hole\\
(Current mass in $M_{\odot}$)&(erg~${\rm s}^{-1}$)&(erg~${\rm s}^{-1}$)&mass ($M_{\odot}$)\\
\hline
$M_{14}$($5\times 10^{14})$&$1.2 \times 10^{46}$&$6.2 \times 10^{43}$&$8.9\times 10^9$\\
$M_{13}$($5\times 10^{13})$&$6.2 \times 10^{44}$&$7.8 \times 10^{42}$&$1.8\times 10^9$\\
$M_{12}$($5\times 10^{12})$&$6.8 \times 10^{44}$&$5.5 \times 10^{42}$&$5.0\times 10^9$\\
$M_{11}$($5\times 10^{11})$&$4.8 \times 10^{43}$&$1.7 \times 10^{41}$&$2.1\times 10^9$\\
\hline
\end{tabular}
}

\textbf{Notes:} The time-averaged supernova power is estimated with a Kroupa IMF (using Eq. 2 of \citealt{2001MNRAS_kroupa}) in which the cold gas mass within $40~{\rm kpc}$ is used as a proxy for the stellar mass. 
 
\label{table:fdbk}
\end{table*}

 For the behavior of gas at large scale, feedback power is more important than the specific feedback implementation/mode -- whether it is driven by black hole accretion or supernovae. 
Keeping that in mind, our model constrains the average feedback power, so as to reproduce the stellar-mass and halo-mass relation at lower redshifts. However, the energy requirement can hint on the feedback mode applicable for a given halo. In order to estimate the typical supernova power, we first consider the cold gas mass within 40 kpc as a proxy for the stellar mass.
 Then we use the Kroupa IMF (using Eq. 2 of \citealt{2001MNRAS_kroupa}) to estimate the number of stars with mass $> 8 ~{\rm M_{\odot}}$ (which explode as supernova) formed at each time, per unit time. Each supernova injects $10^{51}$ ergs of energy in to the halo, in the central region. Therefore the average supernova power 
 can be estimated by
\ba
\label{eq:eqsn}
\frac{\Delta E_{\rm SN}}{\Delta t} = 10^{51}\frac{\Delta N_{\rm SN}}{\Delta t} ({\rm erg~s^{-1}}).
\ea

Table \ref{table:fdbk} shows the feedback power (calculated from average growth of black hole as $\epsilon_{\rm f} [\Delta M_{\rm BH}/\Delta t]{\rm c}^2$ where $\Delta t = $ accumulation duration) and the supernova power estimate (as calculated above) by considering the entire time between $z=6$ and $z=0$. The time-averaged supernova power is more than $2$ orders of magnitude less than the average feedback power required for all the halos. It is well known for clusters at low-redshifts that cooling flows cannot be prevented by feedback from supernovae only. Hence the requirement of AGN feedback is expected for clusters. The predicted feedback powers from observations of X-ray cavities can be as high as $\approx 10^{46}~{\rm ergs^{-1}}$ (\citealt{2008ApJ_birzan}). 
 
The feedback powers for $M_{13}$ and $M_{12}$ are comparable in our model. Initially the seed black hole mass is higher in $M_{13}$ and $M_{12}$ accretes rapidly in the first $1-2~{\rm Gyr}$. In contrast, $M_{13}$ shows a  gentler evolution. Large cavities in groups and galaxies are known to have X-ray luminosities ranging over $7$ orders of magnitude between $10^{38}$ to $10^{45}~{\rm ergs^{-1}}$ (discussed in \citealt{2012NJPh_mcnamara}). There have been only a few observations of elliptical galaxies for X-ray cavities and these observations suggest that the AGN-inflated cavity powers can reach up to $> 10^{43}~{\rm ergs^{-1}}$ (\citealt{2009AIPC_nulsen}). On the other hand, in the  three-dimensional hydrodynamic simulations of a giant elliptical galaxy by \citealt{2012MNRAS_gaspari} the  instantaneous mechanical jet power can be $\gtrsim 10^{44}~{\rm ergs^{-1}}$. Hence, the average feedback energy requirement in our halos seem to be clearly on the higher side.

\subsubsection{Why feedback power is large?}

Compared to the observed jet powers, we have higher feedback power in our cluster run. In fact, the feedback power is very high even in  the smallest halo. This may be partly because many of the smallest halos in the universe are within a cluster/group and the ICM strips significant amount of gas from them. This effect is definitely not captured in our model of isolated galaxies. 

Observations suggest that feedback in cool cluster cores should efficiently compensate for radiative losses without destroying the dense cores. 
However, if we deposit thermal energy isotropically at the center in a small volume, a low-density bubble enclosed by a dense shell is formed. In fact, the dense shell (in which most of the core mass is swept up) can cool very efficiently and enhance cooling losses with such a feedback (e.g., see the top-left panel in Figure 8 of \citealt{2017ApJ...841..133M}). This dense shell
is pushed out by a wind and most of the injected feedback energy goes into moving it against the gravity of the halo, rather than simply balancing cooling 
losses (e.g., see the left panel of Figure \ref{fig:fig09} at 2.5 Gyr). Star formation in our model can only be suppressed if this cold dense shell is moved beyond 40 kpc. 
Adiabatic expansion of the gas may require a large fraction of the feedback energy to be used up in the $PdV$ work done in lifting up the gas from a deep potential well (see also, APPENDIX C of \citealt{2008MNRAS_mccarthy}). 

Observations show that the cluster cores are not isotropically evacuated, 
and anisotropic injection of energy via kinetic jets with small opening angles is more successful. With anisotropic injection most of the energy seems to go in
balancing core cooling (and hence maintaining a reasonable feedback power) rather than uplifting most of the gas in the core (which takes up much larger mechanical power). 
In our 1D simulations we are lifting up all the gas in the center and hence the feedback energy required is larger than in the observed clusters.

\section{Discussion}
\label{sec:disc}

We propose a very simple, generalized 1D model for gas in dark matter halos in which the cosmological growth of the halos and accretion of baryons are taken into account on an average.  We incorporate idealized models for radiative cooling and feedback heating to quantify the energy budget characteristic of each halo (satisfying the abundance matching relation between stellar mass and halo mass). We extend the existing 1D models (\citealt{1978ApJ_parrenod}, \citealt{1997MNRAS_knight}) and emphasize the importance of central energy source, by adding realistic amount of feedback heating proportional to the spherical accretion rate. This enables us to compare the temporally varying state variables of the medium as well as the time-integrated quantities with those of existing models and observations and test the validity of the model which grows halos smoothly over cosmological times. 

The thermodynamics of the gas in clusters and groups is interesting to study, particularly with the advent of future galaxy redshift surveys and the measurement of the kinetic and thermal Sunyaev-Zel'dovich signal (\citealt{2017arXiv_lim}, \citealt{2018ApJ_park}, \citealt{2016PhRvD_schaan}) combined with the existing X-ray observations. In order to test a large number of ICM models and reasonable parametric profiles to compare with new observations, simple semi-analytical models (\citealt{2017ApJ_flender} ) are preferred over large scale hydrodynamic simulations. However, our 1D model, with hydrodynamic evolution of baryons, is computationally less expensive. Moreover, this model can be easily extended to multidimensional simulations of gas in halos without directly evolving dark matter particles. Thus great simplification can be achieved without entirely losing the physics of cosmological accretion of baryons and dark matter. 

Following are the important conclusions from our 1D model:
\begin{itemize}
\item We see the cosmological infall of gas, starting from a small initial halo, undergoing virial shock at radius slightly greater than $r_{200}$, which is fixed by the mass accretion history. But the virial radius is roughly half of the turnaround radius of the falling gas shells. Using a simple profile for the gas reservoir outside the halo, we can ascertain that the average baryon fraction within the halo, in absence of cooling and feedback, is close to the universal value. This provides an attractive setup for multi-dimensional simulations in which the outer boundary is much further out than the virial radius. 
\item For pure radiative cooling, cold mode accretion may dominate for very small halos ($\lesssim 10^{11}~{\rm M}_{\odot}$). This halo mass is a few times smaller than  in \cite{2003MNRAS_birnboimdekel}. So our model incorporates both hot and cold mode accretion in appropriate conditions. In the smallest halo, the gas cools out and loses pressure support fast in the absence of central heating and this often raises the baryon fraction beyond the universal value. On the contrary, the biggest halo retains enough hot gas even in the absence of feedback. This causes it to maintain the universal baryon fraction as the outskirts are hot and pressure-supported while only the gas in the central region cools and falls to the center. The cosmological accretion rate in the cluster-scale halos is  high and a large amount of shock-heated gas continuously joins the halo. However, we revert to the 2D/3D generalisation for studying the detailed physics and survival of cold streams joining the halo in cold-mode. 
\item We tune our feedback (modelled as Bondi-Hoyle-Lyttleton accretion) parameters to obtain a stellar mass-halo mass relation consistent with abundance matching ( \citealt{2010ApJ_moster}).
We use the total cold gas mass within the central $40~{\rm kpc}$ as a proxy for the stellar mass. 
These fiducial runs including cooling and feedback provide realistic estimates of the energy budget in each halo. These fiducial runs lead to the following inferences:
\begin{itemize}
\item The baryon fraction evolution of all the halos show signatures commonly predicted; e.g., smaller halos have majority of baryons missing due to the ejection by feedback and the biggest halos maintain the baryons by intermittent cooling and heating cycles (\citealt{prasad15}). In the former case, the cycles are delayed as it takes a long time for the gas to be recycled, while on cluster scales, the feedback-heated gas remains inside the halo and moves to the center quite fast. 

\item The time-averaged density profile for our cluster-scale halos match well with {\it Chandra} clusters and recent Sunyaev-Zel'dovich-selected clusters followed-up in X-rays (\citealt{2017ApJ_mcdonald}, 
Figure \ref{fig:fig17}). Note that, we are only concerned with the average number density profiles in this work. The feedback efficiencies and the mass accretion history can be tweaked to study different kinds of cluster profiles (cool-core properties). We will revert to this in our future multidimensional version of the model. The interplay of local thermal instability and gravity, which gives the important parameter $t_{\rm cool}/t_{\rm ff}$, can also be modelled only in more than one dimensions (\citealt{choudhury16}).

\item A flattening of the equation of state (or broken power-law as discussed in \citealt{2017ApJ_flender} ) is seen in all the halos that include cooling and heating. Inside the virial shock, $\Gamma$ (where $p \propto \rho^{\Gamma}$) is around $1.1$ while in the core it falls down to around $\approx 0.4-0.6$.  

\item We use a crude estimate of the black hole mass from the idealized Bondi accretion rate (Eddington limited). We see that in groups and clusters the black holes, with a reasonably high seed mass at $z=6$, grow significantly till only around $z=2$. 

\item The supernova power estimated from Kroupa IMF and our simple estimate of the star formation rate, suggest that AGN feedback is relevant for all the halos and absolutely necessary for clusters.
The average feedback power ($\approx 10^{46}$) in $M_{14}$ is comparable to the highest powers observed for many cooling clusters with X-ray cavities and radio lobes. This implies for all our halos, the feedback powers are overestimated. We conclude that in 1D models, a large fraction of the thermal energy gets used up in uplifting the gas. In multidimensional version of our model, AGN jets will uplift gas in its wake, along a specific direction, and the central gas shells are not entirely ejected like in our 1D model. 
 
\end{itemize}
\end{itemize}

However, the general properties of this simple model are consistent with observations and other existing models. It will be extremely efficient and reasonably accurate to survey the physical parameter space using this model in 3D, incorporating smooth cosmological accretion, radiative cooling and feedback heating self-consistently. This may provide an ideal testbed to study the average evolution of the diffuse gas of the CGM, particularly at large radii and outskirts of the halos, which are now observable by current spectroscopic surveys.

\section{Acknowledgement}
PPC is grateful to Max Planck Institute for Astrophysics, Garching, Germany, for the visiting fellowship during her long term visit. GK acknowledges KITP, Santa Barbara and discussions with Anne Thoul. PS acknowledges an India-Israel joint research grant (6-10/2014[IC]).
\bibliographystyle{mn2e}
\bibliography{bibtex}
\label{lastpage}
\end{document}